\documentclass[prb,twocolumn,amsmath,amssymb,nofootinbib]{revtex4-2}
\usepackage{cancel}
\usepackage{xcolor}
\usepackage{graphicx}
\usepackage{dcolumn}
\usepackage{bm}
\usepackage{amsmath}
\usepackage{amsfonts,amssymb,txfonts}
\usepackage{epsfig,color}
\usepackage{graphicx}
\usepackage{hyperref}
\usepackage{float}
\usepackage{comment}
\hypersetup{colorlinks=true,linkcolor=blue,anchorcolor=blue,citecolor=blue,filecolor=blue,urlcolor=blue,bookmarksnumbered=true,pdfview=FitB}
\newcommand{\ve}{\varepsilon}

\newcommand{\bk}{{\bf k}}
\newcommand{\bp}{{\bf p}}

\newcommand{\bq}{{\bf q}}

\newcommand{\kf}{k_{\rm F}}
\newcommand{\ef}{\varepsilon_{\rm F}}
\newcommand{\vf}{\varv_{\rm F}}
\newcommand{\nuf}{\nu_{\rm F}}

\newcommand{\nn}{\nonumber}
\newcommand{\beq}{\begin{equation}}
\newcommand{\eeq}{\end{equation}}
\newcommand{\bea}{\begin{eqnarray}}
\newcommand{\eea}{\end{eqnarray}}
\newcommand{\bse}{\begin{subequations}}
\newcommand{\ese}{\end{subequations}}
\newcommand{\bwt}{\begin{widetext}}
\newcommand{\ewt}{\end{widetext}}

\newcommand{\bu}{{\bf u}}

\newcommand{\ti}{\tau_{\rm i}}

\begin{document}
\preprint{APS/123-QED}

\title{Magnetoconductivity due to electron-electron interaction\\
in a non-Galilean--invariant 
Fermi liquid
}

\author{Tatia Kiliptari}\email{t.kiliptari@ufl.edu} \author{Dmitrii L. Maslov}%
\affiliation{%
Department of Physics, University of Florida, Gainesville, FL 32611-8440, USA 
}%

\date{\today}

\begin{abstract}
The $T^2$-scaling of resistivity with temperature is often viewed as a classic hallmark of a Fermi-liquid (FL) behavior in metals. However, if umklapp scattering is suppressed, this scaling is not universally guaranteed to occur. In this case, the resistivity behavior is influenced by several factors, such as dimensionality (two vs. three), topology (simply- vs. multiply-connected Fermi surfaces), and
(in two dimensions) the shape (convex vs. concave) of the Fermi surface 
(FS).
Specifically for an isotropic spectrum, as well as for a two-dimensional (2D) convex FS, the $T^2$ term is absent, and the first non-zero contribution scales as $T^4\ln T$ in 2D and as $T^4$ in 3D.
In this paper, we study the $T$-dependence of the resistivity, arising from electron-electron interactions, in the presence of a weak magnetic field. We show that, for an isotropic FS in any dimensions and for a convex 2D FS, the $T^2$ term is also absent in both Hall and diagonal components of the magnetoconductivity, which instead scale as $BT^4\ln T$ and $B^2T^4\ln T$, respectively, in 2D and as $BT^4$ and $B^2T^4$ in 3D. The FL-like scaling, i.e., $BT^2$ and $B^2T^2$ of the Hall and diagonal conductivities is recovered for a concave FS in 2D.
Furthermore, we show that, for an isotropic spectrum, magnetoresistance is absent even in the presence of electron-electron interactions. 
Additionally,
we examine the high-temperature limit,
when electron-electron 
scattering prevails over electron-impurity 
one, and show that all the components of the conductivity tensor saturate in this limit at values that are determined by impurity scattering but, in general, differ from the corresponding values at $T=0$.
\end{abstract}

\maketitle

\section{\label{sec:level1}Introduction}
It is well established that electron-electron (\emph{ee}) interactions can influence the resistivity of a non-Galilean invariant Fermi-liquid (FL) metal.
The $T^2$ scaling of resistivity, which is regarded as a characteristic feature of 
FL behavior in metals, can be understood through the Pauli exclusion principle. This principle indicates that at low temperatures, only quasiparticles near the Fermi energy, within a width on the order of $T$, participate in binary collisions. It is important to note, however, that this argument pertains to the inverse quasiparticle relaxation time $1/\tau_\mathrm{ee}$ rather than the resistivity itself. For example, a Galilean-invariant FL has zero resistivity but finite thermal conductivity and viscosity, whose temperature dependences  follow from the $T^2$ scaling of $1/\tau_\mathrm{ee}$. 
This is because in a Galilean-invariant system velocities are proportional to momenta, and conservation of momentum automatically implies conservation of the electric current. Hence, a momentum relaxation mechanism is needed in order to achieve a steady-state current. 

In metals, the presence of a lattice breaks Galilean invariance, allowing for current relaxation through umklapp scattering \cite{landau1936collected} which conserves quasi-momentum up to a reciprocal lattice vector. For umklapp scattering to be permitted, the momentum transfer must be on the order of a reciprocal lattice vector. This requires two conditions to be met: (i) the Fermi surface (FS) must be sufficiently large \cite{abrikosov:book}, and (ii) the interactions must be of sufficiently short range \cite{maslov:2011}. In conventional metals, these conditions are typically satisfied due to the high density of charge carriers and effective screening of Coulomb interactions. As a result, umklapp collisions occur at rates comparable to $1/\tau_\mathrm{ee}$ and $\rho\propto T^2$.

Nevertheless, there are cases where these conditions fail to hold. For example, the first condition 
is breached in systems with low carrier densities, such as degenerate semiconductors, semimetals, and the surface states of three-dimensional topological insulators. The second condition can also be compromised when a metal is tuned to the vicinity of a Pomeranchuk-type quantum phase
transition \cite{maslov:2011}. The interest in conduction mechanisms  due to electron-electron (\emph{ee}) interaction but without umklapps has been rekindled by recent experiments, in which a pronounced $T^2$- behavior of the resistivity was observed in low-density electron systems, when umklapps can be safely ruled out \cite{kamran:2015,Collignon:2018,Lv:2019,Wang_BOS:2020,guo2024fluctuation,Kovalev:2025}.

If umklapp processes are suppressed and the temperature is too low for  electron-phonon interaction to be effective, the primary mechanism for current relaxation becomes electron-impurity \emph{ei} collisions. However, normal \emph{ee} collisions can also influence resistivity under certain conditions. This occurrence depends on the following properties of the Fermi surface (FS): (i) dimensionality (two vs. three), (ii) topology (simply vs. multiply connected FS), and (iii) 
shape  (convex vs. concave) \cite{Gurzhi:1982,gurzhi:1995,maslov:2011,pal:2012b,levitov:2019}. For example, the $T^2$  term is absent
not only for a Galilean-invariant FL but also for any isotropic spectrum both in two and three dimensions.
In two dimensions (2D), the conditions are more restrictive. Namely, the $T^2$ term is absent for a simply connected and convex, yet otherwise arbitrarily anisotropic FS. This occurs because the $T^2$ term originates from collisions between electrons restricted to move along the FS contour. As a result, momentum and energy conservation severely reduce the number of possible scattering channels, such that a convex FS contour behaves similarly to the integrable one-dimensional case, where no relaxation is possible \cite{pal:2012b}.

Whether the $T^2$ is present is determined entirely by a change of the total (group) velocity of two electrons due to a collision (proportional to a change in the electric current):
\begin{equation}\label{eqn:deltav}
\Delta\boldsymbol{\varv}=\boldsymbol{\varv}_\mathbf{k}+\boldsymbol{\varv}_\mathbf{p}-
    \boldsymbol{\varv}_{\mathbf{k-q}}-\boldsymbol{\varv}_{\mathbf{p+q}},
\end{equation}
where $\bk$ and $\bp$ are the initial momenta of electrons and $\bq$ is the momentum transfer.
For example, in a Galilean-invariant system, i.e, a system with parabolic spectrum $\ve_\bk=k^2/2m$, one has $\boldsymbol{\varv}_\bk=\bk/m$, and same for other velocities, such that $\Delta\boldsymbol{\varv}$ vanishes identically due to momentum conservation (in this case, not only the $T^2$ term but all higher-order terms are absent). For a system with isotropic but non-parabolic spectrum, $\Delta\boldsymbol{\varv}$ vanishes if all momenta are taken to lie on the FS. Expanding near the FS, one obtains $T^4\ln T$  and $T^4$ for the leading non-zero term in 2D and 3D, respectively.  Likewise, a 2D convex FS allows only for two scattering channels--the Cooper channel, in which $\bk=-\bp$, and the swap channel, in which $\bp=\bk-\bq$, such that $\Delta\boldsymbol{\varv}=0$ for both channels. Expanding near the Cooper and swap solutions, one again obtains $T^4\ln T$ for  the leading term. A concomitant suppression of the optical conductivity for the same geometries was considered in Refs.~\cite{guo2024fluctuation, rosch:2005,rosch:2006,maslov:2017b,Sharma:2021,Goyal:2023,Li:2023,Esterlis:2021,shi2023loop,shi2024controlled,Gindikin:2024}
both in the FL and non-FL regimes.

The main goal of this paper is to see
if this trend still holds for the 
magnetoconductivity in the presence of a weak magnetic field. In part, our motivation stems from the fact that the hitherto unexplained $T^2$-behavior of the resistivity is often accompanied by anomalous magnetoresistance, which is strong, quasilinear \cite{Lv:2019,kamran_MR:2021}, and, in the case of SrTiO$_3$, almost independent of the mutual orientation of the electric current and magnetic field \cite{kamran_MR:2021}.
\emph{A priori}, the magnetoconductivity probes more detailed characteristics of the spectrum than just the group velocity. For example,  
the magnetoconductivity in the relaxation-time approximation (RTA)
is given by \cite{Ziman:EP} 
\bse
 \bea\label{eqn:sigmaxyimp}
\sigma_{\alpha\neq \beta}(B)&=&-2 e^3\tau^2 B 
\int \frac{d\varepsilon_\mathbf{k}}{(2\pi)^D}(-n'_\bk)
\oint\frac{da_\mathbf{k}}{\varv_\mathbf{k}} \varv_{\bk,\alpha}
u_{\mathbf{k},\beta},\\
 \label{eqn:sigmaxximp}
\delta\sigma_{\alpha\alpha}(B)&=&-2 e^4\tau^3 B^2\int \frac{d\varepsilon_\mathbf{k}}{(2\pi)^D}(-n'_\bk)
\oint\frac{da_\mathbf{k}}{\varv_\bk}
\varv_{\bk,\alpha}w_{\bk,\alpha},
\eea
\ese
where $\tau$ is the relaxation time,  the magnetic field $\mathbf{B}$ is applied along the $z$-axis, the electric field ${\bf E}$ lies in the $xy$ plane, 
 $\alpha=(x,y,z)$, $da_\bk$ is the FS element, $n'_\bk\equiv \partial n(\ve_\bk)/ \partial\ve_\bk$ with $n_\bk\equiv n(\ve_\bk)$ being the Fermi function, and 
\begin{subequations}
\begin{eqnarray} 
    {\bf u}_\mathbf{k}&=&
    \varv_{\bk,x}\frac{\partial \boldsymbol{\varv}_\mathbf{k}}      {\partial k_y}-\varv_{\mathbf{k},y}\frac{\partial 
    \boldsymbol{\varv}_\mathbf{k}}{\partial k_x},
    \label{eqn:u}\\
    {\bf w}_\mathbf{k}&=&
    \varv_{\mathbf{k},x}\frac{\partial {\bf u}
    _\mathbf{k}}{\partial k_y}-\varv_{\mathbf{k},y}\frac{\partial{\bf  u}
    _\mathbf{k}}{\partial k_x}.
    \label{eqn:w}
\end{eqnarray}
\end{subequations}
As we see, unlike for zero field, the magnetoconductivity tensor 
contains higher derivatives of the dispersion, namely, the Hall conductivity in Eq.~\eqref{eqn:sigmaxyimp} contains the components of the (inverse) effective mass tensor, $(1/m)_{\alpha\beta}=\partial \varv_\alpha/\partial k_\beta=\partial^2 \ve_\bk/\partial k_\alpha\partial k_\beta$, while the diagonal magnetocondictivity in Eq.~\eqref{eqn:sigmaxximp} contains the third derivatives of $\ve_\bk$.

In a Galilean-invariant system, \emph{ee} interaction does not lead to current relaxation, even in the presence of a magnetic field. 
This is a consequence of Larmor's theorem \cite{Landau:FT}:
if all particles have the same charge-to-mass ratio, 
the 
forces of mutual Coulomb 
interaction
cancel
each other, 
 thus having no impact on the motion of the center of mass.\footnote{
A related result in the quantum regime is known as Kohn's theorem \cite{Kohn:1961}.} 
In multi-band systems, the condition of equal charge-to-mass ratios is not satisfied.
As a result, \emph{ee} interaction does affect magneto-transport properties of 
these systems,
even if the individual bands are approximated as parabolic (see  Ref.~\cite{levinson:book} and references therein.).
The focus of this paper is the effect of \emph{ee} interaction on magneto-transport properties  of  a single-band, non-Galilean system,  i.e., either an isotropic system with a non-parabolic dispersion of charge carriers or an anisotropic system.
We will show that
the vanishing of $T^2$ terms in all components of the magnetoconductivity tensor
in a non-Galilean-invariant system follows the same rules as in the absence of the magnetic field. 
Namely, such terms vanish for an isotropic but non-parabolic electron energy spectrum, both in 2D and 3D, and for a convex FS in 2D, while the leading terms behave as  $T^4\ln T$ in 2D  and as $T^4$ in 3D.

The remainder of the paper is 
organized as follows. We begin by formulating the problem in terms of the Boltzmann equation (BE) in Sec.~\ref{sec:BE}.
In Sec.~\ref{sec:lowT}, 
we solve the BE perturbatively, first with respect to the magnetic field, and then 
with respect to \emph{ee}
scattering, which is a valid approximation 
for sufficiently low temperatures and 
weak magnetic fields. In Sec.~\ref{sec:iso}
we show that, for  an isotropic but non-parabolic spectrum, the Hall 
and diagonal magnetoconductivity behave as $BT^4\ln T$ and  $B^2T^4\ln T$, respectively, in 2D, and as  $BT^4$ and $B^2T^4$ in 3D. In the same section, we show that \emph{ee} interactions do not give rise to magnetoresistance for the isotropic case.
In Sec.~\ref{sec:convex}, we extend the analysis to a 2D convex and concave FSs. In Sec.~\ref{sec:highT} we discuss the opposite regime of high temperatures, where \emph{ee} contributions to all components of the magnetoconductivity tensor saturate. Our conclusions are given in Sec.~\ref{sec:concl}.

\section{\label{sec:BE} Boltzmann equation: Generalities}
We consider a BE with time-independent and spatially uniform electric  $\mathbf{E}$ and magnetic $ \mathbf{B}$ fields 
\begin{equation}
    -e\;(\mathbf{E}+\boldsymbol{\varv}_\bk\times\mathbf{B})\cdot \frac{\partial f_\mathbf{k}}{\partial \mathbf{k}}
    = -I_{\mathrm{ee}}[f_\mathbf{k}]-I_{\mathrm{ei}}[f_\mathbf{k}],
\end{equation}
Here, $-e$ is the electron charge and $f_\mathbf{k}$ is the distribution function. The collision integrals $I_\mathrm{ee}$ and $I_\mathrm{ei}$ on the right-hand side account for the effects of \emph{ee} and electron-impurity (\emph{ei}) interactions, respectively:
\begin{multline} \label{eqn:Iee}
    I_{\mathrm{ee}}[f_\mathbf{k}]=
\int_{\mathbf{p}}\int_{\mathbf{k}}\int_{\mathbf{k'}}
W_{\mathbf{k},\mathbf{p}\leftrightarrow\mathbf{k'}\mathbf{p'}}
\delta(\varepsilon_\mathbf{p}+\varepsilon_\mathbf{k'}-\varepsilon_\mathbf{k}-\varepsilon_\mathbf{p'})\\ \times
\delta(\mathbf{k}+\mathbf{p}-\mathbf{k'}-\mathbf{p'})\times
[f_\mathbf{k}f_\mathbf{p}(1-f_\mathbf{k'})(1-f_\mathbf{p'})
\\-f_\mathbf{k'}f_\mathbf{p'}(1-f_\mathbf{k})(1-f_\mathbf{p})],
\end{multline}
and
\begin{equation} \label{eqn:ei}
I_{\mathrm{ei}}=\int_\mathbf{k'}
R_{\mathbf{k}\leftrightarrow\mathbf{k'}}
   (f_\mathbf{k}-f_\mathbf{k'})\delta(\varepsilon_\mathbf{k}-\varepsilon_\mathbf{k'}),\end{equation}
where $R_{\bk\leftrightarrow\bk'}$ and $W_{\mathbf{k},\mathbf{p}\leftrightarrow\mathbf{k'}\mathbf{p'}}$ are the corresponding scattering  kernels, and $\int_\mathbf{k}$ is a short-hand notation for $\int d^Dk/(2\pi)^D$ (and the same for other momenta).

For a weak electric field, the driving term $\mathbf{E}\cdot(\partial f_\mathbf{k}/\partial\mathbf{k})$ is reduced to $\boldsymbol{\varv}_\mathbf{k}\cdot\mathbf{E}n'_\mathbf{k}$, where $\boldsymbol{\varv}_\mathbf{k}=\partial \ve_\bk/\partial\bk$ is the electron group velocity. As usual, we define a non-equilibrium part of $f_\bk$ as follows:
\begin{equation}\label{eqn:fk}
\delta f_\mathbf{k}=f_\bk-n_\bk=
n_\mathbf{k}(1-n_\mathbf{k})g_\mathbf{k}=-Tn'_\mathbf{k}g_\mathbf{k}.
\end{equation}
Substituting 
Eq.~(\ref{eqn:fk}) into Eq.~(\ref{eqn:Iee}) and linearizing in $g_\bk$ yields \cite{abrikosov:book}:
\begin{multline}
\label{Ieelin}
 I_\mathrm{ee}[f_\mathbf{k}]=
\int_{\mathbf{p}}\int_{\mathbf{k}}\int_{\mathbf{k'}}
W_{\mathbf{k},\mathbf{p}\leftrightarrow\mathbf{k'}\mathbf{p'}}\\ \times
\delta(\varepsilon_\mathbf{k}+\varepsilon_\mathbf{p}-\varepsilon_\mathbf{k'}-\varepsilon_\mathbf{p'})\times
\delta(\mathbf{k}+\mathbf{p}-\mathbf{k'}-\mathbf{p'})\\ \times
(g_\mathbf{k}+g_\mathbf{p}-g_\mathbf{k'}-g_\mathbf{p'})\times
n_\mathbf{k}n_\mathbf{p}(1-n_\mathbf{k'})(1-n_\mathbf{p'}).
\end{multline}

For concrete results, we will assume that electrons interact via a screened Coulomb potential:
\begin{equation}\label{eqn:Coul}
    U(q)=\frac{2\pi e^2}{q+\kappa}\;
    \text{(in 2D) and}\; U(q)=\frac{4\pi e^2}
    {q^2+\kappa^2}\;\text{(in 3D)}.
\end{equation}
Recalling the Thomas-Fermi expressions for the inverse screening radius,
$\kappa=2\pi e^2\nu_F$ in 2D  and $\kappa^2=4\pi e^2\nu_F$ in 3D with $\nu_F$ being the density of states at the Fermi energy, Eq.~(\ref{eqn:Coul}) can be 
rewritten
as follows:
\begin{equation}\label{eqn:Coul1}
    U(q)=\frac{1}{\nu_F} \frac{\kappa}{q+\kappa}\;
    \text{(in 2D) 
    and}\; U(q)=\frac{1}{\nu_F} \frac{\kappa^2}{q^2+\kappa^2}\;\text{(in 3D)}.
\end{equation}

In the first Born approximation, the \emph{ee} scattering kernel is given by
\bea
W_{\mathbf{k},\mathbf{p}\leftrightarrow\mathbf{k'}\mathbf{p'}}=2\pi U(\bk-\bk')\left[U(\bk-\bk')-\frac 12 U(\bp-\bk')\right].\label{FGR}
\eea
For a weakly screened Coulomb interaction with $\kappa\ll k_{\rm F}$, the second term in the square brackets in Eq.~\eqref{FGR} can be neglected, so that
\bea
W_{\mathbf{k},\mathbf{p}\leftrightarrow\mathbf{k'}\mathbf{p'}}=2\pi U^2(q),\label{stat_scr}
\eea
where $\bq=\bk-\bk'=\bp'-\bp$.

For point-like impurities the \emph{ei} collision integral is reduced to the form:
\begin{equation}\label{eqn:Ieii}
     I_\mathrm{ei}=-\frac{f_\mathbf{k} - \langle f_\bk\rangle}{\tau_\mathrm{i}}.
\end{equation}
On the other hand, the collision integral in the relaxation-time approximation (RTA) reads
\begin{equation}\label{eqn:Iei}
     I_\mathrm{RTA}=-\frac{f_\mathbf{k} - n_\mathbf{k}}{\tau},
\end{equation}
where $\tau$ is a phenomenological relaxation time. The difference between Eqs.~\eqref{eqn:Ieii} and \eqref{eqn:Iei} is important for transport in a non-uniform electric field \cite{Mirlin:1997}; however,  both forms give the same result for the conductivity in a uniform electric field.
In what follows, we will use the RTA form, $I_{\rm ei}=I_{\rm RTA}\vert_{\tau=\ti}$; a more detailed justification of such a replacement in given in  Appendix \ref{sec:RTA}. We will also neglect 
a (usually) weak dependence of $\tau_i$ on $\ve_\bk$.

After these simplifications, the BE is reduced to 
\bea\label{eqn:BE}
     -e(\mathbf{E}\cdot\boldsymbol{\varv}_{\bk})
     n'_\mathbf{k}-e(\boldsymbol{\varv}_\bk
\times\mathbf{B})\cdot\frac{\partial f_\mathbf{k}}{\partial \mathbf{k}}
    =
    -\frac{f_\mathbf{k} - n_\mathbf{k}}{\tau_\mathrm{i}}-I_\mathrm{ee}[f_\mathbf{k}],
    \eea
with $I_{\rm ee}$ given by Eq.~\eqref{Ieelin}. For the remainder of the paper, the magnetic field is assumed to be directed along $z$ axes and weak in the sense that $\omega_\mathrm{c}\tau_\mathrm{i}\ll1$, where $\omega_c=eB/m_c$ is the cyclotron frequency and $m_c$ is the cyclotron mass, and the temperature is assumed to be much lower than the Fermi energy.
\section{\label{sec:lowT}Low temperatures:\\ 
weak electron-electron scattering} 
\subsection{Perturbation theory in electron-electron scattering}
In this section, we examine the low-temperature regime, when 
\emph{ee} collisions are less frequent than \emph{ei} collisions, i.e $1/\tau_\mathrm{ee}\ll1/\tau_\mathrm{i}$, where $1/\tau_\mathrm{ee}$ will be properly defined below. 
Therefore, \emph{ee} scattering can be treated as a correction to 
\emph{ei} one. To begin, we solve Eq.~(\ref{eqn:BE}) with $\mathbf{B}=0$ and $I_\mathrm{ee}=0$, which yields:
\begin{equation}
    g_\mathbf{k}^{(0)}=-\frac{e\tau_\mathrm{i}\mathbf{E}\cdot\boldsymbol{\varv}_\mathbf{k}}{T}.\label{gk0}
\end{equation}
Next, we substitute $g_\mathbf{k}^{(0)}$ back into (\ref{eqn:BE}) and find a linear-in-$B$ correction
\begin{equation}\label{eqn:g1}
     g_\mathbf{k}^{(B)}=\frac{e^2\tau_\mathrm{i}^2
     \boldsymbol{E}\cdot\mathbf{u_\mathbf{k}} }{T}B,
\end{equation}
where $u_\alpha$ is given by Eq.~(\ref{eqn:u}). Performing one more iteration in $B$, we obtain a quadratic-in-$B$ correction
\begin{equation}
g_\mathbf{k}^{(B^2)}=\frac{e^3\tau_\mathrm{i}^3
    \boldsymbol{E}\cdot\mathbf{w}_\mathbf{k}}{T} B^2,
\end{equation}
where $w_\alpha$ is given by Eq.~\eqref{eqn:w}. Finally, we iterate in $I_{\rm ee}$ to obtain the correction due to both $B$ and \emph{ee} scattering:
\begin{equation}\label{eqn:g2}
    g_\mathbf{k}^{({\mathrm{ee}})}(B)=\frac{\tau_\mathrm{i}}{n'_\mathbf{k}T}I_{\mathrm{ee}}[g_\bk^{(0)}+g_\mathbf{k}^{(B)}+g_\mathbf{k}^{(B^2)}].
\end{equation}
The electric current is found as
\bea\label{current}
j_\alpha=-2e\int_\bk \varv_\alpha \delta f_\bk. 
\eea
The leading term in the conductivity is due to impurity scattering:
\bea
\sigma^{(\mathrm{ei})}_{\alpha\beta}=\frac{2e^2\ti }{(2\pi)^2}\int \frac{da_\bk}{\varv_{\bk}}\varv_{\bk,\alpha}\varv_{\bk,\beta}.
\eea

The correction to $\sigma^{\mathrm{ei}}_{\alpha\beta}$ due to \emph{ee} scattering at $B=0$ is found by retaining only the $g_\bk^{(0)}$ term on the right-hand side of Eq.~\eqref{eqn:g2} \cite{maslov:2011,pal:2012b}. Using the symmetry properties of $W_{\mathbf{k},\mathbf{p}\leftrightarrow\mathbf{k'}\mathbf{p'}}$ (see Appendix \ref{app:symmetry}), we obtain

\begin{multline}\label{eqn:s}
\delta\sigma^{(\mathrm{ee},0)}_ {\alpha\beta}=
-\frac{e^2\tau_\mathrm{i}^2}{2T}\int \frac{d^Dq}{(2\pi)^{3D}}\iiint^{\infty}_{-\infty} 
d\omega d\varepsilon_\bk d\varepsilon_\bp \\ 
\times\oint\oint\frac{da_\mathbf{k}}{\varv_\mathbf{k}}\frac{da_\mathbf{p}}{\varv_\mathbf{p}} W_{\mathbf{k},\mathbf{p}\leftrightarrow\mathbf{k'}\mathbf{p'}}
\Delta \varv _\alpha \Delta \varv _\beta\ n(\varepsilon_\mathbf{k}) n(\varepsilon_\mathbf{p})\\ \times
[1-n(\varepsilon_\bk-\omega)][1-n(\varepsilon_\bp+\omega)] \\ \times
\delta(\varepsilon_\mathbf{k}-\varepsilon_{\mathbf{k-q}}-\omega)
\delta(\varepsilon_\mathbf{p}-\varepsilon_{\bp+\bq}+\omega).
\end{multline}
Likewise, retaining the $g_\mathbf{k}^{(B)}$ term in Eq.~\eqref{eqn:g2} yields the correction to the Hall conductivity due to \emph{ee} interaction
\begin{multline}\label{eqn:hallcond}
    \delta \sigma^{(\mathrm{ee},B)}_{\alpha\neq\beta}=
\frac{e^3\tau_\mathrm{i}^3 B}{2T}\int \frac{d^Dq}{(2\pi)^{3D}}\iiint^{\infty}_{-\infty} d\omega d\varepsilon_\bk d\varepsilon_\bp \\ 
     \times
    \oint\oint\frac{da_\mathbf{k}}{\varv_\mathbf{k}}\frac{da_\mathbf{p}}{\varv_\mathbf{p}}
   W_{\mathbf{k},\mathbf{p}\leftrightarrow\mathbf{k'}\mathbf{p'}}
     \Delta \varv
     _\alpha \Delta u
     _\beta\ n(\varepsilon_\mathbf{k}) 
   n(\varepsilon_\mathbf{p})\\  \times
   [1-n(\varepsilon_\bk-\omega)][1-n(\varepsilon_\bp+\omega)] \\ \times
   \delta(\varepsilon_\mathbf{k}-\varepsilon_{\mathbf{k-q}}-\omega)
   \delta(\varepsilon_\mathbf{p}-\varepsilon_{\bp+\bq}+\omega).
\end{multline}
 where 
 \bea \label{Deltau}
 \Delta \bu=\bu_\bk+\bu_\bp-\bu_{\bk-\bq}-\bu_{\bp+\bq},
  \eea
a ${\bf u}$ is given by Eq.~\eqref{eqn:u}, and $\alpha,\beta=(x,y)$.
Finally, retaining the $g_\bk^{(B^2)}$ term in Eq.~\eqref{eqn:g2}, we obtain a correction to the diagonal magnetoconductivity due to \emph{ee} interaction

\begin{multline}\label{eqn:magnetoo}
    \delta \sigma^{(\mathrm{ee},B^2)}_{\alpha\alpha}=
    \frac{e^4\tau_\mathrm{i}^4 B^2}{2T}\int \frac{d^Dq}{(2\pi)^{3D}}\iiint^{\infty}_{-\infty} d\omega d\varepsilon_\bk d\varepsilon_\bp \\ 
     \times
    \oint\oint\frac{da_\mathbf{k}}{\varv_\mathbf{k}}\frac{da_\mathbf{p}}{\varv_\mathbf{p}}
W_{\mathbf{k},\mathbf{p}\leftrightarrow\mathbf{k'}\mathbf{p'}}
     \Delta \varv_\alpha \Delta w_\alpha n(\varepsilon_\mathbf{k}) 
   n(\varepsilon_\mathbf{p})\\  \times
   [1-n(\varepsilon_\bk-\omega)][1-n(\varepsilon_\bp+\omega)] \\ \times
   \delta(\varepsilon_\mathbf{k}-\varepsilon_{\mathbf{k-q}}-\omega)
   \delta(\varepsilon_\mathbf{p}-\varepsilon_{\bp+\bq}+\omega),
\end{multline}
where
 \bea
 \Delta \bf w={\bf w}_\bk+\bf w_\bp-{\bf w}_{\bk-\bq}-{\bf w}_{\bp+\bq}\label{Deltaw},\eea
 and ${\bf w}$ is given by Eq.~\eqref{eqn:w}.

Already the general expressions \eqref{eqn:hallcond} and  \eqref{eqn:magnetoo} reveal the main result of this paper: The same constraints that suppress the \emph{ee} contribution to the conductivity at $B=0$ remain in force at finite $B$ as well. Indeed, the integrands of both Eqs.~\eqref{eqn:hallcond} and  \eqref{eqn:magnetoo} contain 
a factor of $\Delta \varv_\alpha$. Therefore, if $\Delta \varv_\alpha$ vanishes, so do the corrections to the Hall and diagonal magnetoconductivity, regardless
of the behavior of higher derivatives of the dispersion, entering via $\Delta u_\alpha$ and $\Delta w_\alpha$.
 
\subsection{Isotropic but non-parabolic spectrum}\label{sec:iso}
In this section, we consider the case of 
an isotropic but
non-parabolic spectrum, $\ve_\bk=\ve(k)$,  both in 2D and 3D.  
In this case, the group velocity can be 
written
as:
\begin{equation}\label{eqn:iso}
    \boldsymbol{\varv}_\mathbf{k}=\frac{\mathbf{k}}{m(k)},
\end{equation}
where $m(k)=k/(\partial\varepsilon_\bk/\partial k)$ is the density-of-states mass. Then Eqs.~\eqref{eqn:u} and \eqref{eqn:w}  for $u_{\bk,\alpha}$ and $w_{\bk,\alpha}$ are reduced to $u_{\bk,\alpha}=(-k_y\delta_{x\alpha}+k_x\delta_{y\alpha})/m^2(k)$ and 
$w_{\bk,\alpha}=k_\alpha/m^3(k)$. 
If all the momenta are projected onto the FS, i.e., if $|\mathbf{k}|=|\mathbf{p}|=|\mathbf{k-q}|=|\mathbf{p+q}|=k_\mathrm{F}$, 
we get $\Delta\boldsymbol{\varv}=0$, $\Delta \mathbf{u}=0$, and $\Delta \mathbf{w}=0$, and the \emph{ee} contributions to the corresponding components of the magnetoconductivity tensor vanish.  To get a finite result, we need to expand  $\Delta\boldsymbol{\varv}
$, $\Delta \mathbf{u}
$, and $\Delta \mathbf{w}
$ in the vicinity of the FS \cite{Sharma:2021}, such that
$|\mathbf{k}|=k_\mathrm{F}+(\varepsilon_\mathbf{k}-\ef)/\varv_\mathrm{F}$, and similarly for other momenta.  Performing such an expansion, we obtain:
\begin{widetext}
\begin{subequations}
\begin{eqnarray}\label{eqn:varv}
    \Delta \varv_\alpha&=&-\frac{1}{\varv_\mathrm{F}}
    \frac{m'_{\rm F}}{m^2_{\rm F}}
\left[(\varepsilon_\mathbf{k}-\varepsilon_\mathbf{k-q})k_\alpha+
(\varepsilon_\mathbf{p}-\varepsilon_\mathbf{p+q})p_\alpha
+(\varepsilon_\mathbf{k-q}-\varepsilon_\mathbf{p+q})q_\alpha\right]\label{eqn:varv},\\
    \Delta u_\alpha&=&
     \frac{2}{\varv_\mathrm{F}}
     \frac{m'_{\rm F}}{m^3_{\rm F}}
     \left
     \{
     \left[
(\varepsilon_\mathbf{k}-\varepsilon_\mathbf{k-q})k_y+
(\varepsilon_\mathbf{p}-\varepsilon_\mathbf{p+q})p_y
+(\varepsilon_\mathbf{k-q}-\varepsilon_\mathbf{p+q})q_y
\right]\delta_{x\alpha}-
\left[
(\varepsilon_\mathbf{k}-\varepsilon_\mathbf{k-q})k_x+
(\varepsilon_\mathbf{p}-\varepsilon_\mathbf{p+q})p_x
+(\varepsilon_\mathbf{k-q}-\varepsilon_\mathbf{p+q})q_x\right]
\delta_{y\alpha}
\right\},\nn\\
\label{eqn:uu}\\
     \Delta w_\alpha&=&
-\frac{3}{\varv_\mathrm{F}}
     \frac{m'_{\rm F}}{m^4_{\rm F}}
\left[(\varepsilon_\mathbf{k}-\varepsilon_\mathbf{k-q})k_\alpha+
(\varepsilon_\mathbf{p}-\varepsilon_\mathbf{p+q})p_\alpha
+(\varepsilon_\mathbf{k-q}-\varepsilon_\mathbf{p+q})q_\alpha\right],\label{eqn:cc}
\eea
\ese
\end{widetext}
where $m_{\rm F}\equiv m(\kf)$ and $m'_{\rm F}\equiv \partial m(k)/\partial k\vert_{k=\kf}$.  The terms proportional to $q$ in the equations are small for $q\sim\kappa\ll\kf$. Neglecting these terms, we obtain:
\newpage
\begin{subequations}
\begin{eqnarray}\label{eqn:varv}
    \Delta \varv_\alpha&=&-\frac{1}{\varv_\mathrm{F}}
    \frac{m'_{\rm F}}{m^2_{\rm F}}
\left[(\varepsilon_\mathbf{k}-\varepsilon_\mathbf{k-q})k_\alpha+
(\varepsilon_\mathbf{p}-\varepsilon_\mathbf{p+q})p_\alpha
\right]\label{eqn:vvv},\\
    \Delta u_\alpha&=&
     \frac{2}{\varv_\mathrm{F}}
     \frac{m'_{\rm F}}{m^3_{\rm F}}
     \left
     \{
     \left[
(\varepsilon_\mathbf{k}-\varepsilon_\mathbf{k-q})k_y+
(\varepsilon_\mathbf{p}-\varepsilon_\mathbf{p+q})p_y
\right]\delta_{x\alpha}\right.\nn\\
&&\left.-
\left[
(\varepsilon_\mathbf{k}-\varepsilon_\mathbf{k-q})k_x+
(\varepsilon_\mathbf{p}-\varepsilon_\mathbf{p+q})p_x
\right]
\delta_{y\alpha}
\right\},\nn\\
\\
     \Delta w_\alpha&=&-
\frac{3}{\varv_\mathrm{F}}
     \frac{m'_{\rm F}}{m^4_{\rm F}}
\left[(\varepsilon_\mathbf{k}-\varepsilon_\mathbf{k-q})k_\alpha+\label{eqn:www}
(\varepsilon_\mathbf{p}-\varepsilon_\mathbf{p+q})p_\alpha
\right].
\eea
\ese

Using the last results, we first calculate the \emph{ee} contribution to the Hall conductivity $\sigma_{xy}$ of a 2D electron system.
Using  the energy-conserving delta functions in Eq.~(\ref{eqn:hallcond}), we rewrite $\Delta \varv_x$, $\Delta u_y$, and $\Delta w_x$ as 
$\Delta \varv_x=-(m'_{\rm F}/(\varv_\mathrm{F}m^2_{\rm F})\omega(k_x-p_x)$, $\Delta u_y=
-(2m'_{\rm F}/\varv_\mathrm{F}m^3_{\rm F})\omega(k_x-p_x)$, and $\Delta w_x=-(m'_{\rm F}/(\varv_\mathrm{F}m^4_{\rm F})\omega(k_x-p_x)$.
After this step, one can set $\omega=0$ in the delta functions, because $\omega\sim T\ll E_F$.
Substituting $\Delta \varv_x\Delta u_y$ into Eq.~\eqref{eqn:hallcond}, we obtain
\begin{widetext}
        \begin{multline}\label{eqn:hall}
   \delta\sigma^{(\mathrm{ee},B)}_{xy}=
\frac{e^3\tau_\mathrm{i}^3k_\mathrm{F}^2}{T\varv_\mathrm{F}^4}
\frac{{{m'}_{\rm F}^2}}{m^5_{\rm F}}B
\int \frac{d^2q}{(2\pi)^6}
\iiint^{\infty}_{-\infty}  d\omega
d\varepsilon_\mathbf{k} 
d\varepsilon_\mathbf{p}
\int^{2\pi}_0 d\theta_\mathbf{kq}
\int ^{2\pi}_0 d\theta_\mathbf{pq}
W_{\mathbf{k},\mathbf{p}\leftrightarrow\mathbf{k'}\mathbf{p'}}\\
\times \delta(\varepsilon_\mathbf{k}-\varepsilon_\mathbf{k-q}
)
 \delta(\varepsilon_\mathbf{p}-\varepsilon_\mathbf{p+q}
 )
n(\varepsilon_\mathbf{k})n(\varepsilon_\mathbf{p})
[1-n(\varepsilon_\mathbf{k}-\omega)][1-n(\varepsilon_\mathbf{p}+\omega)]
\omega^2(k_x-p_x)^2,
\end{multline}
\end{widetext}
where $\theta_{{\bf n}{\bf m}}$ is the angle between vectors $\mathbf{n}$ and $\mathbf{m}$.
Note that if $\bk\;||\;\bp$, the factor $(k_x-p_x)^2=0$. This implies that out of two possible scattering channels--swap and Cooper--only the Cooper one is operational for the case of forward scattering, considered here.
Next we evaluate the angular integrals assuming again $q\ll \kf$, such that 
the arguments of the delta functions become
$\varepsilon_\mathbf{k}-\varepsilon_\mathbf{k-q}\approx \varv_\mathrm{F}q\cos\theta_{\bk\bq}$ and similarly $\varepsilon_\mathbf{p}-\varepsilon_\mathbf{p+q}\approx-\varv_\mathrm{F}q\cos\theta_{\bp\bq}$. 
 For example, the second delta function is then reduced to 
 $ \delta(\varepsilon_\mathbf{p}-\varepsilon_\mathbf{p+q})=
[\delta(\theta_{\bp\bq}+\pi/2)+\delta(\theta_{\bp\bq}-\pi/2)]/q \vf$. Let $\theta_{\bq\hat x}=\theta$ be the angle between $\mathbf{q}$ and the (arbitrarily chosen) $x$-axis. Then 
$\theta_{\bp\hat x}=\theta_{\bp\bq}+\theta_{\bq\hat x}=\pm \pi/2+\theta$, and the $x$ projection of 
$\bp$ becomes
$p_x=p\cos\theta_{\bp\hat x}\approx k_\mathrm{F}\cos(\pm \pi/2+\theta)=\mp k_\mathrm{F}\sin\theta$; and similarly for 
other the components: 
$k_x=\mp k_\mathrm{F}\sin\theta $, $p_y=\mp k_\mathrm{F}\cos\theta$, and $k_y=\mp k_\mathrm{F}\cos\theta$. Combining all these results, we obtain the expression for the Hall conductivity:
 \begin{widetext}
     \begin{multline} \label{eqn:hall1}
   \delta\sigma^{(\mathrm{ee},B)}_{xy}=
\frac{8\pi e^3\tau_\mathrm{i}^3k^4_F}{(2\pi)^6T\varv_\mathrm{F}^6}
\frac{{{m'}_{\rm F}^2}}{m_\mathrm{F}
^5}B
\int_{T/\vf} \frac{dq}{q}
W_{\mathbf{k},\mathbf{p}\leftrightarrow\mathbf{k'}\mathbf{p'}}
\int d\omega \omega^2
\int d\varepsilon_\mathbf{k} \int d\varepsilon_\mathbf{p}
n(\varepsilon_\mathbf{k})n(\varepsilon_\mathbf{p})
[1-n(\varepsilon_\mathbf{k}-\omega)][1-n(\varepsilon_\mathbf{p}+\omega)],
\end{multline}
 \end{widetext}
where we cut the logarithmic singularity in the integral over $q$ at $q\sim T/\vf$. The integrals  over $\omega$, $\ve_\bk$, and $\ve_\bp$ yield
$8T^5\pi^4/{15}$.  Using the screened Coulomb potential (\ref{eqn:Coul}),
we finally obtain to leading logarithmic accuracy:
\begin{equation}\label{eqn:Hall}
  \delta\sigma^{(\mathrm{ee},B)}_{xy}
=2\sigma^{\mathrm{ei}}
\omega_c\tau_\mathrm{i}\frac{\tau_\mathrm{i}}{\tau_\mathrm{ee}},
\end{equation}
where 
\begin{equation}\label{eqn:drude}
    \sigma^{\mathrm{ei}}=\frac{e^2}{2\pi}
   m_{\rm F}\vf^2 \tau_\mathrm{i}
\end{equation}
is the residual conductivity and 
\begin{equation}\label{eqn:tauee}
    \frac{1}{\tau_\mathrm{ee}}=\frac{2\pi^3}{15}
 \left(\frac{\partial\ln m}{\partial\ln k}\right)^2\Big\vert_{k=\kf}
  \frac{T^4}
{m_\mathrm{F}^3\varv_\mathrm{F}^6}\ln\frac{\varv_\mathrm{F}\kappa}{T}
\end{equation}
is the effective \emph{ee} scattering rate. The latter was defined in such a way that the correction to the zero-field conductivity due to \emph{ee} interaction, Eq.~\eqref{eqn:s}, conforms to the Matthiessen rule: $\delta\sigma^{(\rm ee,0)}=-\sigma_{\rm ei} \tau_{\rm i}/\tau_{{\rm ee}}$. Note that $1/\tau_{\rm ee}=0$ if $m$ does not depend on $k$, i.e., if the system is Galilean-invariant.
Equation (\ref{eqn:Hall}) 
shows that for a 2D isotropic spectrum, the correction to the Hall conductivity due to \emph{ee} interactions behaves as $BT^4\ln T$, which is the same $T$-dependence as of the correction to the zero-field conductivity \cite{Sharma:2021}.
To calculate the diagonal magnetoconductivity, we use Eq.~(\ref{eqn:magnetoo}) along with Eqs.~(\ref{eqn:varv}) and (\ref{eqn:www}). 
Repeating similar steps, we obtain 
\begin{equation}\label{eqn:Magneto}
    \delta\sigma_{xx}^{(\mathrm{ee},B^2)}=3\sigma^{\mathrm{ei}}
(\omega_c\tau_\mathrm{i})^2\frac{\tau_\mathrm{i}}{\tau_\mathrm{ee}},
\end{equation}
which behaves as $B^2 T^4\ln T$ in 2D.

In the 3D case, following similar steps as in Eqs.~(\ref{eqn:hallcond}) and (\ref{eqn:magnetoo}), we obtain Eqs.~(\ref{eqn:Hall}) and (\ref{eqn:Magneto}), respectively. However, $1/\tau_\mathrm{ee}$ in 3D is proportional just to $T^4$, 
without a logarithmic factor.

It is well known that there is no magnetoresistance for an isotropic electron spectrum  in the presence of impurity scattering alone, because the $B$-dependences of $\sigma_{xx}$ and $\sigma_{xy}$ cancel each other on inverting the conductivity tensor. To determine whether \emph{ee} gives rise to magnetoresistance,
we expand $\rho_{xx}$ up to quadratic order in $
B$:
\begin{equation}\label{eqn:rho}
    \rho_{xx}=\frac{1}{\sigma^{\mathrm{\mathrm{ei}}}}\left( 1-\frac{\delta\sigma_{xx}(B)}{\sigma^{\mathrm{ei}}}
    -\frac{\delta\sigma_{xy}^2(B)}{\left(\sigma^{\mathrm{ei}}\right)^2}
    \right),
\end{equation}
where $\delta\sigma_{xx}=
\sigma^{\mathrm{ei}}\left(1-\omega^2_c\tau^2_{\rm i}\right)
+\delta\sigma^{(\rm ee,B^2)}_{xx}
$ and 
$\delta\sigma_{xy}
=-\sigma^{\mathrm{ei}}\omega_c\tau_{\rm i}
+\delta\sigma^{(\rm ee,B)}_{xy}$.
Subsituting Eqs.~\eqref{eqn:Hall} and \eqref{eqn:Magneto} into Eq.~\eqref{eqn:rho}, we find that the magnetic field dependence cancels out and $\rho_{xx}=1/\sigma_{\rm i}$, which is the same as in the absence of \emph{ee} interactions.

The results derived above 
are valid in the limit $T\ll\varv_\mathrm{F}\kappa$. 
If the ratio $\kappa/\kf$ is very small, there also exists an intermediate range of temperatures, $\vf\kappa\ll T\ll \ef$. In this range, electrons remain degenerate, but their behavior is no
longer described by the FL theory. For example, the quasiparticle scattering rate in this temperature range scales as $T$ \cite{agd:1963}, rather than as $T^2$, because electrons are scattered by plasmon modes in the equipartition regime.
The correction to the zero-field conductivity of a 2D Dirac system in this temperature range has recently been shown to scale as $T^2$ \cite{Kovalev:2025}. Although Ref.~\cite{Kovalev:2025} considered a statically screened interaction, it can be shown that 
taking into account dynamic screening does not change the result (see Appendix \ref{app:dyn}). Accordingly, the components of the magnetoconductivity tensor also scale as $T^2$ in this range: The corresponding results are obtained by replacing $1/\tau_{\rm ee}$ in Eqs.~\eqref{eqn:Hall} and \eqref{eqn:Magneto}
with
\begin{equation}\label{eqn:tau-dyn}
\frac{1}{\tau_\mathrm{ee}}=
\frac{\pi}{12}
 \left(\frac{\partial\ln m}{\partial\ln k}\right)^2\Big\vert_{k=\kf}
 \left(\frac{\kappa}{\kf}\right)^2
  \frac{T^2}
{m_\mathrm{F}\varv_\mathrm{F}^2}. 
\end{equation}
In contrast to the FL $T^2$ scaling, however, the temperature dependence of the conductivity for $\vf\kappa\ll T\ll \ef$ is not universal, but depends on the dimensionality of the system: In 3D, the corresponding scattering rate scales as $T$.

It needs to be stressed that the $T^4$ and $T^2$ scaling forms represent only corrections to the residual conductivity and apply only for $T$ below certain temperature $T_{\rm i}$, at which the \emph{ee} and \emph{ei} scattering rates become comparable. For $T\gg T_i$, the conductivity saturates at a $T$-independent value, which is again determined only by impurities \cite{Gurzhi:1968,levinson:book,maslov:2011,pal:2012b,Sharma:2021}. This high-$T$ regime is discussed in detail in Sec.~\ref{sec:highT}.  

    
\subsection{ 
Anisotropic Fermi surface in 2D}
\label{sec:convex}
Now, we turn to the case of a generic anisotropic FS in 2D. 
As stated in Sec.~\ref{sec:level1}, the
$T^2$ contribution to the zero-field 
residual conductivity vanishes for the case of a convex FS, and the leading $T$-dependent term is $T^4\ln T$. On the other hand, the $T^2$ correction is nonzero for a concave FS. Now we will analyze the $T$-dependent corrections to magnetoconductivity.

To see why the $T^2$ correction vanishes for a convex FS, we recall that this correction comes from the scattering processes in which both the initial ($\bk$ and $\bp$) and final ($\bk-\bq$ and $\bp+\bq$) momenta lie on the FS. 
 In 2D, this condition 
 is satisfied only by the Cooper and swap channels, where $\bp=-\bk$ and $\bp=\bk-\bq$, respectively \cite{gurzhi:1995,maslov:2011,pal:2012b}. However, these channels do not relax the current in the absence of the magnetic field because  $\Delta\boldsymbol{\varv}$ in Eq.~\eqref{eqn:deltav} vanishes, and the $T^2$ term in Eq.~\eqref{eqn:s} is absent. Likewise, the vectors $\Delta{\bf u}$ in Eq.~\eqref{Deltau} and 
$\Delta{\bf w}$ in Eq.~\eqref{Deltaw}
also vanish, and there are no $T^2$ corrections to either  Hall or diagonal magnetoconductivities.
To obtain a finite result, one must to expand the vectors $\Delta\boldsymbol{\varv}$, $\Delta{\bf u}$, and $\Delta{\bf w}$ near the Cooper and swap channels.
For $\Delta\boldsymbol{\varv}$, such an expansion gives the current relaxation rate which behaves as $\max\left\{\Omega^4\ln|\Omega|,T^4\ln T\right\}$, where $\Omega$ is the frequency of an oscillatory electric field,    i.e., in the same way as for an isotropic but non-parabolic electron spectrum \cite{Li:2023}.
Here, we will also need expansions of $\Delta{\bf u}$ and $\Delta{\bf w}$, which we demonstrate for the Cooper channel as an example. Expanding momenta close to the swap solution $\mathbf{p}_0=-\bk_0$, as $\mathbf{k}=\mathbf{k}_0+\delta\mathbf{k}$ and 
$\mathbf{p}=\mathbf{p}_0+\delta\mathbf{p}=
-\mathbf{k}_0+\delta\mathbf{p}$,
we have:
\bse
\bea
    \Delta 
    \boldsymbol{\varv}_{\rm C} &=&([\delta\mathbf{k}+\delta\mathbf{p})\cdot
\boldsymbol{\nabla}][
\boldsymbol{\varv}_{\bk_0}-
\boldsymbol{\varv}_{\bk_0-\bq}]
,\label{v:expand}\\
    \Delta 
    {\bf u}_{\rm C}&=&[(\delta\mathbf{k}+\delta\mathbf{p})\cdot
\boldsymbol{\nabla}]
[{\bf u}_{\bk_0}-\bf u_{\bk_0-\bq}]
\label{u:expand},\\
 \Delta 
    {\bf w}_{\rm C}&=&[(\delta\mathbf{k}+\delta\mathbf{p})\cdot
\boldsymbol{\nabla}]
[{\bf w}_{\bk_0}-\bf w_{\bk_0-\bq}].
\label{w:expand}
\eea
\ese
For the Hall conductivity, we obtain:
\begin{widetext}
\begin{multline}
  \delta\sigma^{\mathrm{ee},B}_{xy}=
-\frac{e^3\tau^3_{\rm i} B}{2T(2\pi)^6}
\int d^2q
\int d^2\delta k \int d^2\delta p \int d\omega
W_{\mathbf{k},\mathbf{p}\leftrightarrow \bk'\bp'}
\delta(\varepsilon_\mathbf{k}-\varepsilon_\mathbf{k-q}-\omega)
 \delta(\varepsilon_\mathbf{p}-\varepsilon_\mathbf{p+q}+\omega)\\
\times n(\varepsilon_\mathbf{k})n(\varepsilon_\mathbf{p})
[1-n(\varepsilon_\mathbf{k}-\omega)][1-n(\varepsilon_
{\bp}+\omega)]
\Delta\varv_{{\rm C},x}\Delta u_{{\rm C},y}.
\label{Hallconve}
\end{multline}
\end{widetext}
Although the integrals 
cannot be calculated explicitly for
a generic FS, the $T$-dependence of the result can still be extracted. To this end, we employ the condition of small momentum transfers ($q\ll\kf$) and further expand 
$\mathbf{a}\equiv\boldsymbol{\varv}_{\bk_0-\bq}-\boldsymbol{\varv}_{\bk_0}\approx-(\mathbf{q}\cdot\boldsymbol{\nabla})\boldsymbol{\varv}_{\bk_0}$ and $\mathbf
{\mathbf{b}}
\equiv\mathbf{u}_{\bk_0-\bq}-\mathbf{u}_{\bk_0}\approx-(\mathbf{q}\cdot\boldsymbol{\nabla})\mathbf{u}_{\bk_0}$. Also,
the energy differences are expanded as
$\varepsilon_{\mathbf{k}_0+\delta\bk}-\varepsilon_{\mathbf{k}_0+\delta\bk-\mathbf{q}}
={-\bf a}\cdot\delta\mathbf{k}$ 
and
$\varepsilon_{\mathbf{p}_0+\delta\bp}-\varepsilon_{\mathbf{p}_0+\delta\bp+\mathbf{q}}
=
\varepsilon_{-\mathbf{k}_0+\delta\bp}-
\varepsilon_{-\mathbf{k}_0
+\delta\bp+\bq}
{=\bf a}
\cdot\delta\mathbf{p}$. Substituting the above expansions into Eq.~\eqref{Hallconve} yields:
\begin{widetext}
\begin{multline}
  \delta\sigma^{(\mathrm{ee},B)}_{xy}=
-\frac{e^3\tau^3_{\rm i} B}{2T(2\pi)^6}
\int d^2q
\int d^2\delta k \int d^2\delta p \int d\omega
W_{\mathbf{k},\mathbf{p}\leftrightarrow \bk',\bp'}
\delta(
{\bf a}\cdot\delta\mathbf{k}+\omega) \delta(
{\bf a}\cdot\delta\mathbf{p}+\omega)\\
\times
n(\boldsymbol{\varv}_{\bk_0}\cdot\delta\bk)
n(\boldsymbol{\varv}_{-\bk_0}\cdot\delta\bp)
[1-n(\boldsymbol{\varv}_{\bk_0}\cdot\delta\bk-\omega)][1-n(\boldsymbol{\varv}_{-\bk_0}\cdot\delta\bp+\omega)]
\Delta\varv_{{\rm C},x}\Delta u_{{\rm C},{y}}.
\label{Hallconvex}
\end{multline}
\end{widetext}
The delta functions in Eq.~(\ref{Hallconvex}) reduce the 2D integrals
over $\delta k$ and $\delta p$ to one-dimensional integrals along
the straight lines:
\begin{equation}
   \delta\bk\cdot{\bf a}=-\omega\;
   \text{ and}\; \delta\bp\cdot
   {\bf a}=-\omega.\;\;
\end{equation}
It is convenient to choose $\delta k_x$ and 
$\delta p_x$ as independent integration
variables and exclude $\delta k_y$  and $\delta p_y$ via
\begin{subequations}
\begin{eqnarray} 
    \delta k_y=\label{eqn:deltaky}
    -\frac{\omega}{a_y}-\delta k_x
    \frac{a_x}{a_y},
   \\
   \delta p_y=\label{eqn:deltapy}
    -\frac{\omega}{a_y}-\delta p_x
    \frac{a_x}{a_y}
    .\label{eqn:c}
\end{eqnarray}
\end{subequations}
The Pauli principle (imposed by the Fermi functions functions) and
energy conservation (imposed by the delta-functions), effectively confine
$\delta k_x$ and $\delta p_x$ to the
interval of width proportional to $T$:
\begin{eqnarray} 
  -T/a\lesssim
   \delta k_x,\,\delta p_x
   \lesssim T/a,
   \label{eqn:c}
\end{eqnarray}
where $a\equiv |{\bf a}|$. At the same time, typical energy transfers are also limited by temperature: $|\omega|\sim T$.
Substituting Eqs. (\ref{eqn:deltaky}) and (\ref{eqn:deltapy}) into Eqs. (\ref{v:expand}) and (\ref{u:expand}), 
we obtain near the Cooper solution: 
\begin{subequations}
\begin{eqnarray} 
\Delta \boldsymbol{\varv}_{\rm C}
&=&
-\left[
\delta k_x \left( \partial_{k_x} - \frac{a_x}{a_y} \partial_{k_y} \right) 
+\delta p_y \left(\partial_{k_y}-\frac{a_y}{a_x}
\partial_{k_x}  \right) 
\right. 
\nonumber \\
&&
\left. 
-\frac{\omega }{a_x} \partial_{k_x} - \frac{\omega}{a_y} \partial_{k_y}
\right] \mathbf{a},
   \\
   \Delta \mathbf{u}_{\rm C} &=& 
   -\left[
\delta k_x \left( \partial_{k_x} - \frac{a_x}{a_y} \partial_{k_y} \right) 
+\delta p_y \left(\partial_{k_y}-\frac{a_y}{a_x}
\partial_{k_x}  \right) 
\right. 
\nonumber \\
&&
\left. 
-\frac{\omega }{a_x} \partial_{k_x} - \frac{\omega}{a_y} \partial_{k_y}
\right] \mathbf{b}.
\end{eqnarray}
\end{subequations}
Therefore, $\Delta\boldsymbol{\varv}_{\rm C},\,\Delta\bu_{\rm C}\propto |\omega|\sim  T$.
With all the constraints having been resolved, Eq. (\ref{Hallconvex}) is
reduced to
\begin{multline}
\delta\sigma^{(\mathrm{ee},B)}_{xy}
\sim
-\frac{e^3\tau^3 B}{
T
}
\int d^2q W_{\mathbf{k},\mathbf{p}\leftrightarrow \bk',\bp'}
\frac{1}{
a_y^2}\\\times
\int^T_{-T} d\omega
\int_{
-T/a}^{T/a}
 d\delta k_x 
 \int_{-T/a}^{T/a}
d\delta p_x 
\Delta\varv_{{\rm C},x}\Delta u_{{\rm C},y}.
\end{multline}
Now we can power-count the result. The integrals over $\delta k_x$, $\delta p_x$, and $\omega$ give factor of $T$ each, while another factor of $T^2$ comes from the product $\Delta\varv_{{\rm C},x}\Delta u_{{\rm C},y}$. This leaves us with an overall factor of $T^4$. The integral over $q$ is log-divergent at $q\to 0$ because $a\propto q$. Cutting off the divergence at $T/\vf$, we obtain $\delta\sigma^{(\rm ee,B)}_{xy}\propto BT^4\ln T$, which is the same scaling as in the isotropic case. The contribution from the swap channel leads to the same result. 

The same argument works for the diagonal magnetoconductivity, which contains a factor of $\Delta w$ 
instead of $\Delta u$ but, since $\Delta w\propto T$, we obtain again $\delta\sigma^{(\rm ee,B^2)}_{xy}\propto B^2T^4\ln T$.

For a concave FS in 2D, there are more than two solutions $\bk$ and $\bp$ for the initial momenta for a given $\mathbf{q}$. While some of these solutions still belong to the Cooper and swap channels, 
the additional solutions allow for current relaxation. Consequently, $\delta\sigma^{(\rm ee,B)}_{xy}\propto B T
^2
$ and $\delta\sigma^{(\rm ee,B^2)}_{xy}\propto B^2T^2$.

We 
emphasize
that 
for a convex Fermi surface,
the prefactor of the $T^4\ln T$ term 
is of order unity in terms of the Coulomb coupling constant $r_s\sim\kappa/k_\mathrm{F}$, 
since the electron charge $e$ enters the result only 
logarithmically, as an upper cutoff of the integral over $q$ (cf. Eq.~\eqref{eqn:hall1} for the isotropic case).\footnote{The same logarithmic dependence on $e$ occurs also in other properties af 2D electron system,  such as the quasiparticle lifetime \cite{giuliani:1982}, thermal conductivity \cite{lyakhov:2003}, and viscosity \cite{novikov:2006}.}
In contrast, for a concave Fermi surface, the logarithmic singularity in the momentum integral is suppressed by a $q^2$ factor arising from $\left(\Delta \boldsymbol{\varv}\right)^2$, leading the integral to be dominated  by the region $q\sim \kappa$.
Consequently,   the $T^2$ term acquires a prefactor $\sim r_s^2 \ln(1/r_s)$. 
 The total result for a concave FS is thus a combination of two contributions, a $T^4\ln T$ term and a $\sim r_s^2 \ln(1/r_s)T^2$ term, 
 which can, in principle, become comparable when  $r_s\ll 1$.


\section{\label{sec:highT}
High-temperatures:\\ 
strong electron-electron scattering}
\subsection{General results}
Up to this point, our analysis has concentrated on the low-temperature limit, where the \emph{ee} contribution to resistivity 
is a correction to the \emph{ei} one. 
This regime holds up to a temperature $T_{\rm i}$, at which the \emph{ee} and \emph{ei} scattering times become comparable. For $T\gg T_{\rm i}$, frequent but momentum-conserving \emph{ee} collisions establish local equilibrium in the electron system; however, they cannot fix the value of the drift velocity. In the absence of \emph{ei} scattering, electrons as a whole would be still accelerated by the electric field. The role of \emph{ei} scattering is to provide a balancing frictional force, such that a steady state is achieved. The steady-state current in this regime is controlled entirely by impurities (assuming that electron-phonon scattering can still be neglected), and thus the resistivity saturates at some $T$-independent value, which, in general, is different from the residual one \cite{Gurzhi:1968,levinson:book} (for an isotropic system, the low- and high-$T$ saturation values coincide \cite{Sharma:2021}). Presumably, such a saturation was observed in ultraclean samples of aluminum \cite{Chiang:1966} (although it was attributed to momentum-conserving electron-phonon rather than electron-electron scattering \cite{Gurzhi:1968}.) 

A method of calculating the resistivity in the high-temperature regime, based on the spectral decomposition of the collision integral, was developed in Refs.~\cite{maslov:2011,pal:2012b}. Here, we apply this method to magnetoconductivity.
The \emph{ee} collision integral in Eq.~(\ref{eqn:BE}) can be considered as a linear operator $\hat{I}_{\mathrm{ee}}$, acting on the non-equilibrium part of the distribution function $f_\mathbf{k}^{(1)}\equiv f_\mathbf{k}-n_\mathbf{k}$
\begin{equation}
\hat I_\mathrm{ee}[f_\mathbf{k}^{(1)}]=\sum_\mathbf{k'} 
\hat I_\mathrm{ee}(\mathbf{k},\mathbf{k}') f_\mathbf{k'}^{(1)}.
\end{equation}
The non-Hermitian operator $I_{\mathrm{ee}}$ can be represented
in terms of its left ($\langle\tilde{\Phi}^\lambda_\gamma|$) and right ($|\Phi^\lambda_\gamma\rangle$) eigenstates
as
\begin{equation}
    \hat I_\mathrm{ee}[f_\mathbf{k}^{(1)}]=\frac{1}{\tau_\mathrm{ee}^\star}
\sum_{\lambda,\gamma} \lambda 
 |\Phi^\lambda_\gamma \rangle \langle\tilde{\Phi}^\lambda_\gamma|,\label{ortho}
\end{equation}
where $\lambda$ is the corresponding eigenvalue, $\gamma=x,y,z$, and $\tau_\mathrm{ee}^\star$
 is the effective \emph{ee} scattering time, which sets an overall magnitude of the collision integral. The right
and left eigenstates constitute an orthonormal basis:
\begin{equation}\label{eqn:orth}
    \langle \Tilde{\Phi}^\lambda_{\gamma }|\Phi^{\lambda'}_{\gamma' } \rangle=
    \frac{1}{\mathcal{V}}\sum_\mathbf{k} \Tilde{\Phi}^{\lambda}_{\mathbf{k},\gamma}
    \Phi^{\lambda'}_{\mathbf{k},\gamma'}=
\delta_{\lambda,\lambda'}\delta_{\gamma\gamma'},
\end{equation}
where $\Phi^\lambda_{\bk,\gamma}=\langle\bk|\Phi^\lambda_\gamma\rangle$, $\Tilde{\Phi}^\lambda_{\bk,\gamma}=\langle\Tilde{\Phi}_{\bk,\gamma}|\bk\rangle$
and  $\mathcal{V}$ is the system volume in the $D$-dimensional
space.
A general solution of Eq.~(\ref{eqn:BE}) can be expanded
over this complete basis as
\begin{equation}\label{eqn:f}
    f_\mathbf{k}^{(1)}=\sum_\lambda 
    \mathbf{A}^\lambda
    \cdot
    \boldsymbol{\Phi}^\lambda_\mathbf{k},
\end{equation}
where $\boldsymbol{\Phi}^\lambda_\bk=
\left(\Phi^\lambda_{\bk,x},\,\Phi^\lambda_{\bk,y},\Phi^\lambda_{\bk,z}\right)$
and
$
\mathbf{
A
}
^\lambda=
\mathbf{
A
}
^\lambda_{0}+
\mathbf{
A
}
^\lambda_{1}
+
\mathbf{
A
}
^\lambda_{2}
\dots$ forms a series in powers of the magnetic field, such that ${\bf A}^\lambda_{n}\propto B^n$.
The ${\bf A}_{0}^\lambda$ term corresponds to $B=0$. To find this term, we substitute the series \eqref{eqn:f} into Eq.~\eqref{eqn:BE} with $B=0$, and project the result onto $|\boldsymbol{\Tilde{\Phi}}^\lambda\rangle$ to obtain 
\begin{multline}
\sum_\bk\Tilde{\Phi}^\lambda_{\bk,\rho}e(\boldsymbol{\varv}_{\mathbf{k}} \cdot \mathbf{E} )n'_\bk=
    \frac{1}{\tau_\mathrm{i}} \sum_{\bk,\gamma,\lambda'}\Tilde{\Phi}^\lambda_{\bk,\rho} A^{\lambda'}_\gamma\Phi^{\lambda'}_{\bk,\gamma}\\
   + \frac{1}{\tau_\mathrm{ee}^\star}
    \sum_{\bk,\bk'}\sum_{\gamma,\delta}
    \sum_{\lambda',\lambda''}
    \Tilde{\Phi}^\lambda_{\bk,\rho}
    \Phi^{\lambda'}_{\bk,\gamma}\lambda'
\Tilde{\Phi}^{\lambda'}_{\bk',\gamma}A^{\lambda''}_\delta
    \Phi^{\lambda''}_{\bk',\delta}.
\end{multline}
Using Eq.~(\ref{eqn:orth}) we arrive at:
\begin{equation}
\mathbf{A}^\lambda_{0}=
   \left( \frac{1}{\tau_\mathrm{i}}+
   \frac{\lambda}{\tau_\mathrm{ee}^\star}
   \right)^{-1}
\sum_\bk\boldsymbol{\Tilde{\Phi}}^\lambda_{\bk}e(\boldsymbol{\varv}_{\mathbf{k}} \cdot \mathbf{E} )n'_\bk.
\end{equation}
In
the limit of $1/\tau_\mathrm{ee}^\star\rightarrow\infty$,
only the zero-mode ($\lambda=0$) contribution survives. Therefore, 
\begin{equation}\label{eqn:A0}
\mathbf{A}_0^{0}=e\tau_\mathrm{i}\sum_\bk\boldsymbol{\Tilde{\Phi}}^0_{\bk}(\boldsymbol{\varv}_{\mathbf{k}} \cdot \mathbf{E} )n'_\bk.
\end{equation}
The right and left zero modes are given by \cite{pal:2012b}
\bea
\Tilde{\Phi}^{0}_{\mathbf{k},\gamma}=\tilde{C}_\gamma k_\gamma,
\;\Phi^{0}_{\mathbf{k},\gamma}=
-C_\gamma k_\gamma n'_\bk,\label{Phi0}
\eea
where no summation over $\gamma$ is implied, $\tilde{C}_\gamma$ and $C_\gamma$ are normalized by the condition 
\begin{equation}\label{eqn:norm}
C_\gamma
\Tilde{C}_\gamma
=
[\nuf\langle k_\gamma^2 \rangle]^{-1},
\end{equation}
with  $\nuf$ being the density of states at the Fermi energy,  and  
\begin{equation}\label{eqn:aver}
    \langle F \rangle\equiv\frac{1}{\nuf\mathcal{V}}
    \sum_\mathbf{k}F(\mathbf{k})(-n'_\mathbf{k}).
\end{equation}

Accordingly, the high-temperature limit of the conductivity at $B=0$ reads \cite{pal:2012b}:
 \begin{equation}
\sigma_{\alpha\beta}|_{T\rightarrow\infty}=
    2e^2\tau_{\rm i}\nu_\mathrm{F}
    \sum_{\gamma}
    \frac{\langle \varv_\alpha k_\gamma \rangle}{\langle k_\gamma^2\rangle}
    \langle k_\gamma\varv_\beta \rangle.\label{B=0_highT}
 \end{equation}
For comparison, the residual conductivity at $T=0$ is given by
 \begin{equation}
     \sigma_{\alpha\beta}|_{T\rightarrow 0}=
    2e^2\tau_{\rm i}\nuf
    \langle \varv_\alpha \varv_\beta \rangle.
 \end{equation}

To find a correction to the Hall conductivity, 
we need to iterate Eq.~\eqref{eqn:BE} in the Lorentz-force term once. Keeping again only the zero-mode contribution, we find:
\begin{equation}\label{eqn:A1}
   \mathbf{A}
   ^{0}_1= e\tau_\mathrm{i} \sum_\bk \Tilde{\boldsymbol{\Phi}}^0_\bk(\boldsymbol{\varv}_\mathbf{k}\times\mathbf{B})\cdot
    \frac{\partial}{\partial\mathbf{k}}\left[
    \mathbf
    {A}^{0}_0\cdot
\boldsymbol{\Phi}^0_\mathbf{k}\right].
\end{equation}
Inserting the last result into Eq.~\eqref{eqn:f}, 
we obtain for the corresponding non-equilibrium part of the distribution function
 \begin{multline}\label{eqn:f1}
    f^{(1)}_\mathbf{k}=\sum_{\gamma}A^{0}_{1\gamma}
   \Phi^0_\gamma 
    (\mathbf{k})=
    \frac{e^2\tau_\mathrm{i}^2}{\mathcal{V}^2}\sum_{\mathbf{k'},\mathbf{k''}} 
    \sum_{\gamma,\delta,\beta} C_\gamma k_\gamma n'_\mathbf{k'} \Tilde{C}_\gamma k'_\gamma \\
    \times (\boldsymbol{\varv}_\mathbf{k'}\times\mathbf{B})_\delta n'_\mathbf{k'}
    \Tilde{C}_\delta k''_\delta \varv_{\mathbf{k''},\beta}
    n'_\mathbf{k''} C_\delta E_\beta.
\end{multline}
Taking into account Eqs. (\ref{eqn:norm}) and (\ref{eqn:aver}), we finally obtain:
\begin{equation}\label{Hallhigh}
    \sigma_{\alpha\beta}(B)\Big|_{T\rightarrow \infty}
    =2e^3\tau_\mathrm{i}^2\nuf\sum_{\gamma,\delta}
    \frac{\langle \varv_\alpha k_\gamma \rangle}{\langle k_\gamma^2\rangle}
    \langle k_\gamma (\boldsymbol{\varv} \times {\bf B})_\delta \rangle
    \frac{\langle k_\delta \varv_\beta \rangle}{\langle k_\delta^2\rangle}.
\end{equation}
Iterating in the Lorentz-force one more time, we obtain the diagonal magnetoconductivity

\begin{multline}\label{eqn:maghigh}
   \delta\sigma_{\alpha\alpha}(B)|_{T\rightarrow \infty}\equiv \sigma_{\alpha\alpha}(B)|_{T\to\infty}-\sigma_{\alpha\alpha}(0)|_{T\to\infty}\\
    =2e^4\tau_\mathrm{i}^3 \nuf \\ \times\sum_{\gamma,\delta,\rho}
    \frac{\langle \varv_\alpha k_\gamma\rangle}{\langle k_\gamma^2\rangle}
    \frac{\langle k_\gamma ({\boldsymbol{\varv}}\times {\bf B})_\delta \rangle
    \langle k_\delta({\boldsymbol{\varv}}\times {\bf B})_\rho \rangle}{\langle k_\delta^2\rangle}
       \frac{\langle  k_{\rho} \varv_\alpha \rangle}{\langle k_{\rho}^2\rangle},
\end{multline}  
where $\sigma_{\alpha\alpha}(0)|_{T\to\infty}$ is the diagonal element of Eq.~\eqref{B=0_highT}.

In  the opposite limit of low temperatures, Eqs.~\eqref{eqn:sigmaxyimp} and \eqref{eqn:sigmaxximp} give the standard expressions:
\bse
\bea
    \sigma_{\alpha\beta}(B)|_{T\rightarrow 0}&=&-
    2e^3\tau_\mathrm{i}^2\nuf B
    \langle \varv_\alpha u_\beta \rangle\label{Halllow},\\
\delta\sigma_{\alpha\alpha}(B)|_{T\rightarrow 0}&=&-
    2e^4\tau_\mathrm{i}^3 \nuf B^2
    \langle \varv_\alpha w_\alpha \rangle\label{diaglow},
\eea
\ese
where $u_\alpha$ and $w_\alpha$ are given by equations (\ref{eqn:u}) and (\ref{eqn:w}) respectively.

 It is important to note that the saturation occurs for any dimensionality and shape of the FS; that is, regardless of whether the temperature dependence of a particular component of  the conductivity tensor  begins at low temperatures with a $T^2$ term (like for a concave FS), or with a $T^4\ln T$ term (like for a convex FS), it will eventually saturate at higher temperatures. In practice, however, other scattering mechanisms, such as electron-phonon interaction, may obscure the conductivity saturation.


We now return briefly to the behavior of the conductivity in the intermediate range of temperatures, $ \vf\kappa\ll T\ll\ef$, discussed at the end of Sec.~\ref{sec:iso}. The $T^2$-scaling of the conductivity in this regime can be observed only
if the saturation temperature, $T_{\rm i}$, is much larger than $\vf\kappa$. Then, the temperature dependence of the conductivity starts as $T^4\ln T$, crosses over to $T^2$ at $T\sim \vf\kappa$, and finally, saturates at the $T$-independent value for $T\gg T_\mathrm{i}$. For an isotropic system, the high-temperature saturation value coincides with the residual conductivity (this implies that the conductivity exhibits a minimum at $T\sim T_{\rm i}$ \cite{Sharma:2021}).  Therefore, a $T^2$ behavior is confined to the interval $\vf\kappa\ll T\ll T_{\rm i}\ll \ef$.
To estimate $T_{\rm i}$, we equate $\tau_{\rm ee}$ from Eq.~\eqref{eqn:tau-dyn} to $\tau_{\rm i}$, which yields
$T_{\rm i}\sim (\kf/\kappa)(\ef/\ti)^{1/2}$.
Therefore, the condition $\kappa \vf\ll T_{\rm i}$ is equivalent to
\begin{eqnarray}
  \frac{\kappa}{k_\mathrm{F}}\ll
    \left(\frac
    {1}{\ef\ti}
    \right)^{1/4}\ll 1,
\end{eqnarray}
which is more restrictive than $\kappa\ll \kf$.
\subsection{Specific examples}
The low- and high-temperature limits of all the conductivity components differ in the way that the conductivity is averaged over the FS.
Naturally, the two limits align for an isotropic dispersion, which implies that the zero-field conductivity exhibits a minimum at $T\sim T_{\rm i}$ \cite{Sharma:2021}. Likewise, the Hall and diagonal magnetoconductivities are also expected to have minima at $T\sim T_{\rm i}$. As anisotropy increases, the low- and high-temperature limits starts to differ. A detailed temperature dependence of $\sigma_{\alpha\beta}$ can be obtained only via numerical solution of the Boltzmann equation, which is beyond the scope of this work. What we can readily calculate 
though is the ratio of the low-T/high-T limits, because 
(at least for pointlike impurities) it 
is determined entirely by the FS geometry. In the remainder of this section, we present the results of such a calculation.

As an example, we consider the tight-binding model for a 2D square lattice
with the energy dispersion
 \begin{equation} 
     \varepsilon=4(t+t')-2t(\cos(k_xa)+\cos(k_ya))-4t'\cos(k_xa)\cos(k_ya), \label{eqn:tight}
 \end{equation}
where $t$ and $t'$ denote the hopping amplitudes between nearest
and next-to-nearest-neighbor sites, respectively, and $a$ is the lattice constant. For simplicity, we will set $t=1$ and $a=1$ in what follows.

For both low- and nearly-full fillings, the dispersion in Eq.~(\ref{eqn:tight}) becomes isotropic, and the low- and high-temperature limits of $\sigma_{\alpha\beta}$ coincide.
First, we consider the $t'=0$ case (dashed lines in the figures).
Figure \ref{img:1} shows the ratio of low- and high-temperature conductivities in zero magnetic field. As expected, the ratio is close to 1 for low filling ($\ve_{\rm F}\ll 4t$) and reaches about 2 near half-filling ($\ve_{\rm F}\approx 4t$). 
That $\sigma_0\vert_{T=0}>\sigma_0\vert_{T\to\infty}$ can be understood by recalling  that \emph{ee} interaction in our case reduces the conductivity. Although 
a decrease in the conductivity saturates for $T\gg T_{\rm i}$ (cf. Sec.~\ref{sec:highT}), the high-$T$ limit of $\sigma_0$ is still lower that its $T=0$ value.
[Due to 
particle-hole symmetry 
at $t'=0$, the behavior is the same in the interval from
half-filling at $4t$ to full-filling at $8t$.]
Figure~\ref{img:2} shows the same ratio for the diagonal magnetoconductivity.
The low-$T$/high-$T$ ratio for the diagonal magnetoconductivity behaves qualitatively similar to the zero-field case, except for that the 
enhancement of this ratio near half-filling
is much more pronounced,
reaching a factor of $
17.5
$ at 
$\ef=3.98t$.

\begin{figure}[h!]
\centering
\includegraphics[scale=0.5]{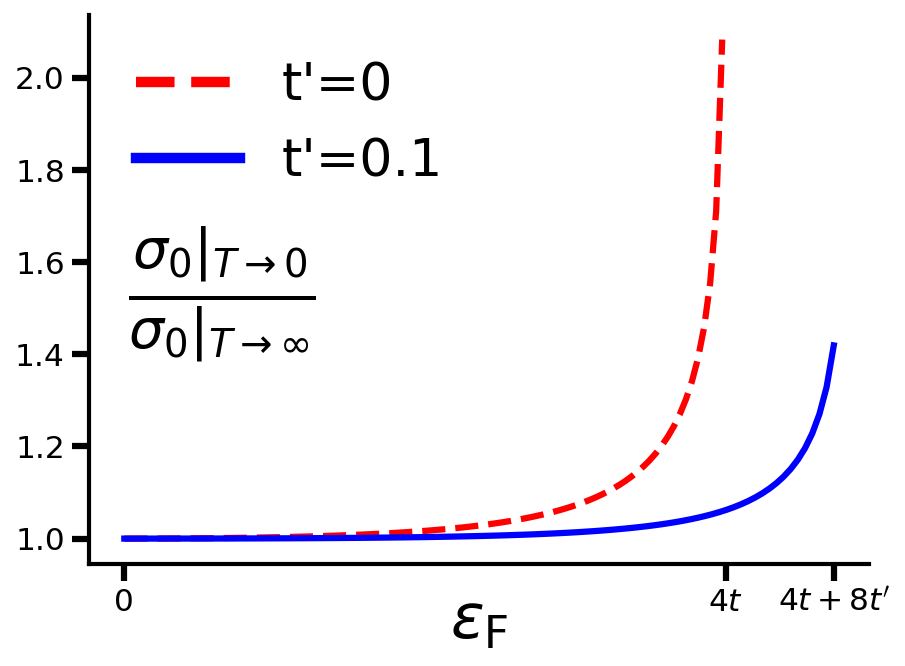}
\caption{
The ratio of the low- and high-temperature limits of the zero-field conductivity for the dispersion in
Eq.~\eqref{eqn:tight} 
with $t'=0$ (dashed line), and $t'=0.1$ (solid line).
Here, $\sigma_0\equiv\sigma_{xx}(B=0)=\sigma_{yy}(B=0)$. Note that $\ve_{\rm F}=4t$ corresponds to half-filling, 
at which point the Fermi surface is 
maximally
anisotropic, 
while $\ve_{\rm F}=4t+8t'$ corresponds to a Van Hove 
singularity.}
\label{img:1}
\end{figure}

\begin{figure}[h!]
\centering
\includegraphics[scale=0.5]{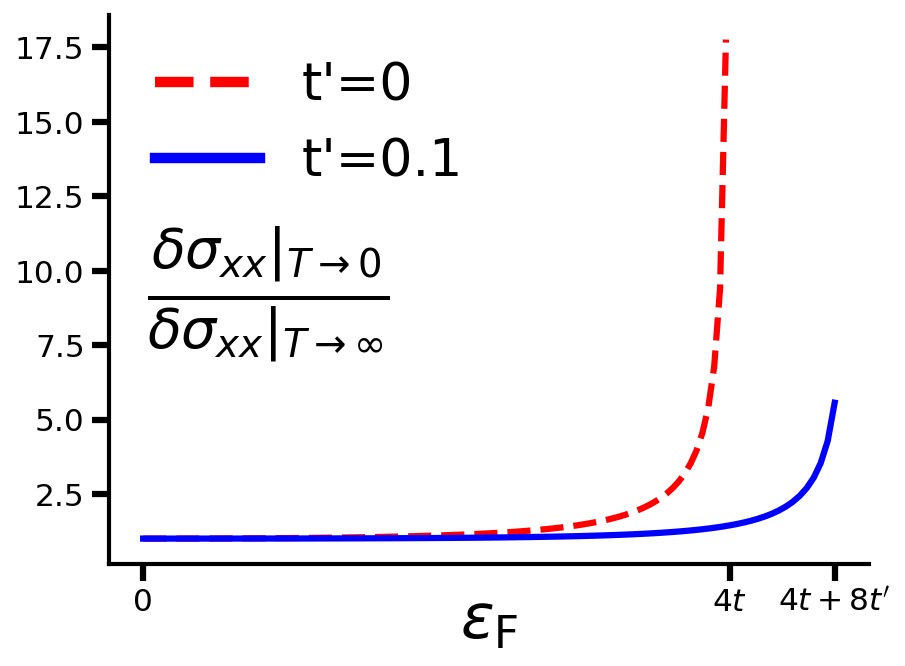}
\caption{
The ratio of the low- and high-temperature limits of diagonal magnetoconductivity for the dispersion in
Eq.~\eqref{eqn:tight} 
with $t'=0$ (dashed line), and $t'=0.1$ (solid line).
}
\label{img:2}
\end{figure}
The Hall conductivity, shown in Fig.~~\ref{img:3}, behaves in the opposite way: the high-$T$ limit of $\sigma_{xy}(B)$ is significantly larger than the low-$T$ one; near half-filling,   $\sigma_{xy}(B)|_{T\to 0}/\sigma_{xy}(B)|_{T\to \infty}\approx 
0.3
$.

This difference in behavior arises because, in contrast to the zero-field conductivity and diagonal magnetoconductivity, the Hall conductivity of a particle-hole symmetric system
vanishes both at zero and half fillings. The same is true both for $\sigma_{xy}|_{T\to 0}$ and $\sigma_{xy}|_{T\to\infty}$, and the key point is which one vanishes more rapidly. Therefore, low-$T$/high-T ratio is very sensitive to the details of the bandstructure.
To examine this sensitivity, we include next-nearest-neighbor hopping, which breaks particle-hole symmetry. As a result, the behavior near 
around half-filling
becomes less sensitive to the exact position of the Fermi energy. For $t'=0.1$ (solid line in Fig.~\ref{img:3}), the low-$T$/high-$T$ ratio for $\sigma_{xy}$ is much closer to unity than for $t'=0$.
The zero-field and diagonal magnetoconductivity are also affected by broken particle-hole symmetry (solid lines in Figs.~\ref{img:1} and \ref{img:2}).

\begin{figure}[h!]
\centering
\includegraphics[scale=0.5]{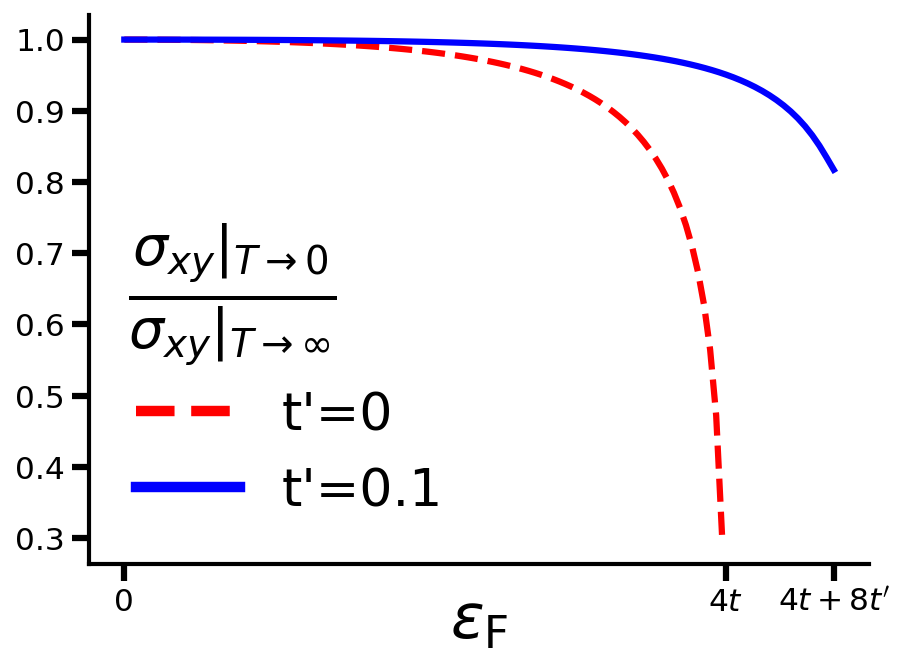}
\caption{
The ratio of the low- and high-temperature limits of the Hall conductivity for the dispersion in
Eq.~\eqref{eqn:tight} 
with $t'=0$ (dashed line), and $t'=0.1$ (solid line).}
\label{img:3}
\end{figure}

\section{Conclusions}
\label{sec:concl}
In this study, we have explored the Hall and diagonal magnetoconductivities arising from electron-electron (\emph{ee}) interactions in a non-Galilean-invariant Fermi liquid. We showed that, at lowest temperatures, a $BT^2$ correction to the Hall conductivity $\sigma_{xy}$ and 
$B^2 T^2$ correction to the diagonal magnetoconductivity $\sigma_{xx}$
are absent for an isotropic but non-parabolic spectrum, both in 2D and 3D.
The leading non-zero contributions 
in this case behave as $\sigma_{xy}\propto BT^4\ln T$ and $\sigma_{xx}\propto B^2T^4\ln T$ 
in 2D
in 2D 
and as $\sigma_{xy}\propto BT^4$ and $\sigma_{xx}\propto B^2T^4$ in 3D.
Although the components of the magnetoconductivity tensor depend on both $B$ and $T$, the diagonal magnetoresistance is absent for an isotropic system, as is also the case without \emph{ee} interactions.
The same scaling behavior, i.e, $BT^4\ln T$ for the Hall and $B^2T^4\ln T$ for the diagonal magnetoconductivity is also found for a convex surface in 2D, while the $BT^2$ and $B^2T^2$ scaling forms are recovered for a concave Fermi surface. 
We also analyzed the high-temperature regime, in which electron-electron scattering dominates over electron-impurity one, leading to a saturation  in both zero-field  and field-dependent corrections due to \emph{ee} interaction. For an isotropic system, the conductivities 
in the high- and low-temperature limits coincide.
 
\acknowledgements
 We are grateful to A. Hall for critically reading the manuscript. This work was supported by the National Science Foundation (NSF)  via grant  DMR-2224000. 
 D. L. M. also acknowledges the hospitality of the Kavli Institute for Theoretical Physics, Santa Barbara, supported by the NSF grants PHY-1748958 and PHY-2309135, and support
 from the Simons Foundation Targeted Grant 920184 to the William I. Fine Theoretical Physics Institute, University of Minnesota.
\section{DATA AVAILABILITY}
The data that support the findings of this article are openly available \cite{kiliptari2025data}.

\onecolumngrid
\appendix
\section{Relaxation-time approximation for impurity scattering}
\label{sec:RTA}
The \emph{ei} collision integral for point-like impurities-function 
is given by Eq.~\eqref{eqn:Ieii}, which we copy below for the reader's convenience
\begin{equation}
    I_\mathrm{ei}=
    \frac{f_\mathbf{k}-\langle f\rangle}{\tau_\mathrm{i}}.
\end{equation} 
Therefore, the solution of the  BE,
\begin{equation}
    -e(\boldsymbol{\varv}\cdot\mathbf{E})\frac{\partial\langle f\rangle}{\partial\ve_\bk}=\frac{\langle f\rangle-f_\mathbf{k}}{\tau_\mathrm{i}},
    \end{equation}
 is defined only up to an arbitrary function of energy, $\langle f\rangle$. This does not affect the conductivity, as the $\langle f\rangle$ drops out from the current, but the distribution function itself remains
undetermined. To fix this, we introduce a phenomenological collision integral of the relaxation-time approximation form
\begin{equation}
I_{\rm RTA}=\frac{n_\mathbf{k}-f_\mathbf{k}}{\tau},
\end{equation}
where $n_\mathbf{k}$ is the equilibrium distribution function.
Note that the integral of $I_{\rm RTA}$ over the directions of $\bk$ at fixed energy is non-zero. 
Therefore, such a term cannot come from potential scattering, because the
latter conserves the number of particles at given energy. At the same time $I_{\rm RTA}$ leads to energy dissipation, because $\int_\bk \ve_\bk I_{\rm RTA}\neq 0$. Now the BE reads
\begin{equation}\label{BErta}
    -e(\boldsymbol{\varv}\cdot\mathbf{E})n'_\bk=\frac{\langle f\rangle-f_\mathbf{k}}{\tau_\mathrm{i}}+\frac{n_\mathbf{k}-f_\mathbf{k}}{\tau},
    \end{equation}
where $\partial\langle f\rangle/\partial\ve_\bk$ was replaced by $n'_\bk$, because now $f$ is expanded
around $n_\bk$, rather than around $\langle f \rangle$. Solving Eq. (\ref{BErta}), we obtain
\begin{equation}
    f_\mathbf{k}=\frac{1}{\frac{1}{\tau_\mathrm{i}}+\frac{1}{\tau}}
    \left[ e(\boldsymbol{\varv}\cdot\mathbf{E})n'_\mathbf{k}+ 
    \frac{\langle f\rangle}{\tau_\mathrm{i}}+\frac{n_\mathbf{k}}{\tau}\right].
\end{equation}
Averaging the last expression over angles,
we find that $\langle f\rangle=n_\mathbf{k}$. Now the solution is unique
\begin{equation}
    f_\mathbf{k}=n_\mathbf{k}+\frac{1}{\frac{1}{\tau_\mathrm{i}}+\frac{1}{\tau}}
     e(\boldsymbol{\varv}\cdot\mathbf{E})n'_\mathbf{k}.
\end{equation}
After this step, we can safely set $1/\tau=0$, which reproduces Eqs.~\eqref{eqn:fk} and \eqref{gk0}.
\section{Symmetrized form of the electron-electron contribution to the conductivity}
\label{app:symmetry}
We assume that our system has both time-reversal and inversion symmetries. In this case, the scattering probability  satisfies the microreversibility condition \cite{sturman:1984,levinson:book}
\begin{equation}
W_{\mathbf{k},\mathbf{p}\leftrightarrow\mathbf{k'}\mathbf{p'}}=
W_{\mathbf{k'},\mathbf{p'}\leftrightarrow\mathbf{k}\mathbf{p}}
.\label{first}
\end{equation}
Next, indistinguishability of electrons implies that 
\begin{equation}
W_{\mathbf{k},\mathbf{p}\leftrightarrow\mathbf{k}^{\prime },\mathbf{p}^{\prime }}=W_{%
\mathbf{k},\mathbf{p}\leftrightarrow\mathbf{p}^{\prime },\mathbf{k}^{\prime }}=W_{\mathbf{p%
},\mathbf{k}\leftrightarrow\mathbf{k}^{\prime },\mathbf{p}^{\prime }}=W_{\mathbf{p},%
\mathbf{k}\leftrightarrow\mathbf{p}^{\prime },\mathbf{k}^{\prime }}. \label{ind}
\end{equation}
Finally, combining the last two properties, we obtain the third one 
\begin{equation}
W_{\mathbf{kp\leftrightarrow k}^{\prime }\mathbf{p}^{\prime }}
=W_{\mathbf{p}^{\prime
}\mathbf{k}^{\prime }\leftrightarrow\mathbf{pk}}.\label{third}
\end{equation}
The first iteration of the BE with respect to magnetic field gives the following form of the correction to the Hall conductivity due to \emph{ee} interactions
\bea
\delta\sigma^{(\mathrm{ee},B)}_{\alpha\beta}&\propto&\int_{\bk,\bp,\bk',\bp'}W_{\mathbf{k},\mathbf{p}\leftrightarrow\mathbf{k'}\mathbf{p'}}
\varv_{\mathbf{k},\alpha}
    [u_{\mathbf{k},\beta}+u_{\mathbf{p},\beta}
    -u_{\mathbf{k'},\beta}-
    u_{\mathbf{p'},\beta}]\mathcal{D}_{\bk,\bp,\bk',\bp'}
   \equiv S, \label{eqn:apa2}
\eea
where $\mathcal{D}_{\bk,\bp,\bk',\bp'}\equiv
    \delta(\bk'+\bp'-\bk-\bp)\delta(\ve_{\bk'}+\ve_{\bp'}-\ve_{\bk}-\ve_{\bp}) n(\ve_\bk) n(\ve_\bp)\left[1-n(\ve_{\bk'})\right]\left[1-n(\ve_{\bp'})\right]$.
Relabeling ($\mathbf{k}\leftrightarrow\mathbf{p}$) and 
$\mathbf{k'}\leftrightarrow\mathbf{p'}$), and using Eq.~\eqref{ind}, we obtain:
\begin{multline}\label{eqn:apa3}
S=  \int_{\bk,\bp,\bk',\bp'} W_{\mathbf{p},\mathbf{k}\to\mathbf{k'}\mathbf{p'}}
    \varv_{\mathbf{p},\alpha}
    [u_{\mathbf{k},\beta}+u_{\mathbf{p},\beta}
    -u_{\mathbf{k'},\beta}-u_{\mathbf{p'},\beta}]\mathcal{D}_{\bk,\bp,\bk',\bp'}=\int_{\bk,\bp,\bk',\bp'}
W_{\mathbf{k},\mathbf{p}\leftrightarrow\mathbf{k'}\mathbf{p'}}
    \varv_{\mathbf{p},\alpha}
    [u_{\mathbf{k},\beta}+u_{\mathbf{p},\beta}
    -u_{\mathbf{k'},\beta}-u_{\mathbf{p'},\beta}]\mathcal{D}_{\bk,\bp,\bk',\bp'}.
\end{multline}
Next, we interchange $\mathbf{k}\leftrightarrow\mathbf{k'}$ and
$\mathbf{p}\leftrightarrow\mathbf{p'}$, and apply Eq.~\eqref{first}, to obtain:
\begin{multline} \label{eqn:apa4}
 S= \int_{\bk,\bp,\bk',\bp'}  W_{\mathbf{k'},\mathbf{p'}\to\mathbf{k}\mathbf{p}}
    \varv_{\mathbf{k'},\alpha}
    [u_{\mathbf{k'},\beta}+u_{\mathbf{p'},\beta}
    -u_{\mathbf{k},\beta}-u_{\mathbf{p},\beta}]\mathcal{D}_{\bk,\bp,\bk',\bp'}=
   - \int_{\bk,\bp,\bk',\bp'} W_{\mathbf{k},\mathbf{p}\leftrightarrow\mathbf{k'}\mathbf{p'}}
    \varv_{\mathbf{k'},\alpha}
    [u_{\mathbf{k},\beta}+u_{\mathbf{p},\beta}
    -u_{\mathbf{k'},\beta}-u_{\mathbf{p'},\beta}]\mathcal{D}_{\bk,\bp,\bk',\bp'}.
\end{multline}
Finally, we interchange $\mathbf{k}\leftrightarrow\mathbf{p'}$ and 
$\mathbf{p}\leftrightarrow\mathbf{k'}$, and apply Eq.~\eqref{third}, to arrive at
\begin{multline} \label{eqn:apa5}
S=  \int_{\bk,\bp,\bk',\bp'}  
  W_{\mathbf{p'},\mathbf{k'}\to\mathbf{p}\mathbf{k}}
    \varv_{\mathbf{p'},\alpha}
    [u_{\mathbf{p'},\beta}+u_{\mathbf{k'},\beta}
    -u_{\mathbf{p},\beta}-u_{\mathbf{k},\beta}]\mathcal{D}_{\bk,\bp,\bk',\bp'}=
   -  \int_{\bk,\bp,\bk',\bp'}W_{\mathbf{k},\mathbf{p}\leftrightarrow\mathbf{k'}\mathbf{p'}}
    \varv_{\mathbf{p'},\alpha}
    [u_{\mathbf{k},\beta}+u_{\mathbf{p},\beta}
    -u_{\mathbf{k'},\beta}-u_{\mathbf{p'},\beta}]\mathcal{D}_{\bk,\bp,\bk',\bp'}.
\end{multline}
Adding up Eqs. (\ref{eqn:apa2}
- \ref{eqn:apa5}), we obtain: 
\begin{eqnarray}
    S&=&\frac{1}{4}  \int_{\bk,\bp,\bk',\bp'}W_{\mathbf{k},\mathbf{p}\leftrightarrow\mathbf{k'}\mathbf{p'}}
    [\varv_{\mathbf{k},\alpha}+\varv_{\mathbf{p},\alpha}
    -\varv_{\mathbf{k'},\alpha}-\varv_{\mathbf{p'},\alpha}]
    [u_{\mathbf{k},\beta}+u_{\mathbf{p},\beta}
    -u_{\mathbf{k'},\beta}-u_{\mathbf{p'},\beta}]\mathcal{D}_{\bk,\bp,\bk',\bp'}\nn\\
    &=&\frac{1}{4}  \int_{\bk,\bp,\bk',\bp'}W_{\mathbf{k},\mathbf{p}\leftrightarrow\mathbf{k'}\mathbf{p'}} \Delta \varv_\alpha \Delta u_\beta \mathcal{D}_{\bk,\bp,\bk',\bp'},
\end{eqnarray}
which is the form announced in Eq.~\eqref{eqn:hallcond} of the main text. A symmetrized form of the diagonal magnetoconductivity in Eq.~\eqref{eqn:magnetoo} is derived along the same lines.

\section{
Correction to the conductivity due to dynamically screened Coulomb interaction
}
\label{app:dyn}
We now turn to the intermediate range of temperatures, $\vf\kappa\ll T\ll \ef$, where, \emph{a priory}, one needs to take into dynamic, rather than static, screening of the Coulomb interaction. We consider the case of zero-field conductivity and deduce the effective current-relaxation time from this quantity. In the isotropic case, the same time enters all the components of the magnetoconductivity tensor. 

The only change compared to case of a statically screened interaction is that the scattering probability is now expressed via the dynamically screened one as \cite{lyakhov:2003}
\bea
W_{\mathbf{k},\mathbf{p}\leftrightarrow\mathbf{k'}\mathbf{p'}}=2\pi |U(q,\omega)|^2,\label{stat_scr}
\eea
where $\bq=\bk-\bk'=\bp'-\bp$, and $\omega=\ve_{\bk}-\ve_{\bk-\bq}=\ve+{\bp+\bq}-\ve_\bp$.  In the random-phase approximation,
\begin{equation}
    U(q,\omega)=
    \frac{U_0(q)}{1-U_0(q)\Pi(q,\omega)}
\end{equation}
where $U_0(q)$ is the bare Coulomb potential and $\Pi(q,\omega)$
is the polarization bubble. In 2D and for $|\omega|/\vf\leq q\ll\kf$,
\bea
\Pi(q,\omega)=-\nu_{\rm F}\left(1+\frac{i\omega}{\sqrt{\vf^2q^2-\omega^2}}\right).
\eea
Accordingly, 
\begin{equation}\label{eqn:CoulDyn}|
    U(q,\omega)|^2=\frac{1}{\nu_{\rm F}^2}
    \frac{\kappa^2(q^2\varv_\mathrm{F}^2-\omega^2)}{(q+\kappa)^2(q^2\varv_\mathrm{F}^2-\omega^2)+\kappa^2\omega^2}.
\end{equation}
Following the same steps as in the main text, we arrive at following form of the correction to the zero-field conductivity
\begin{multline}
   \delta\sigma^{(\mathrm{ee},0)}_{xx}=-
\frac{\pi e^2\tau_\mathrm{i}^2k_\mathrm{F}^2}{
T\varv_\mathrm{F}^4}
\frac{{{m'}_{\rm F}^2}}{m^4_{\rm F}}
\int \frac{d^2q}{(2\pi)^6}
\int^{\infty}_{-\infty}  d\omega \omega^2 
|U(\bq,\omega)|^2
\int^{2\pi}_0 d\theta_\mathbf{kq}
\int ^{2\pi}_0 d\theta_\mathbf{pq}
\delta(\omega-\varv_\mathrm{F}q\cos\theta_{\bk\bq}
)
\delta(\omega-\varv_\mathrm{F}q\cos\theta_{\bp\bq}
)
(k_x-p_x)^2
 \\
\times\int d\varepsilon_\mathbf{k}
\int d\varepsilon_\mathbf{p}
n(\varepsilon_\mathbf{k})n(\varepsilon_\mathbf{p})
[1-n(\varepsilon_\mathbf{k}-\omega)][1-n(\varepsilon_\mathbf{p}+\omega)].
\end{multline}
The integral in the last line yields
\begin{equation}
    \int d\varepsilon_\mathbf{k}
\int d\varepsilon_\mathbf{p}
n(\varepsilon_\mathbf{k})n(\varepsilon_\mathbf{p})
[1-n(\varepsilon_\mathbf{k}-\omega)][1-n(\varepsilon_\mathbf{p}+\omega)]=\frac{1}{2}
\frac{\omega^2
}{\sinh^2{(\omega/2T)}}.
\end{equation}
In contrast to the statically screened case, one cannot neglect $\omega$ in the delta-functions yet. 

Next, we approximate $k_x
=k_\mathrm{F}\cos(\theta_{\bk\mathbf{x}})=  k_\mathrm{F}\cos(\theta_{\bk\bq}+\theta_{\bq{\bf x}})$, and and similarly, $p_x
=p_\mathrm{F}\cos(\theta_{\bp\mathbf{x}})=  k_\mathrm{F}\cos(\theta_{\bp\bq}+\theta)$, where $\theta=\theta_{\bq\mathbf{x}}$.
Then the integrals over $\theta_{\bk\bq}$ and $\theta_{\bp\bq}$ yield
\begin{equation}
\int^{2\pi}_0 d\theta_\mathbf{kq}
\int ^{2\pi}_0 d\theta_\mathbf{pq}
\delta(\omega-\varv_\mathrm{F}q\cos\theta_{\bk\bq}
)
\delta(\omega-\varv_\mathrm{F}q\cos\theta_{\bp\bq}
)
(k_x-p_x)^2=
\frac{8k_\mathrm{F}^2\sin^2\theta}{\varv_\mathrm{F}^2q^2}\Theta(\varv_\mathrm{F}q-\omega)
\end{equation}
Here, $\Theta$ denotes the Heaviside step function. As a result of this simplification, we arrive at the following expression for the interaction-induced Hall conductivity correction:
\begin{equation}\label{eqn:hall-dyn}
     \delta\sigma^{(\mathrm{ee},0)}_{xx}=-
\frac{
8\pi^2k_\mathrm{F}^4e^2\tau_\mathrm{i}^2}{(2\pi)^6T\varv_\mathrm{F}^6\nu_{\rm F}^2}
\frac{{{m'}_{\rm F}^2}}{m^4_{\rm F}}
\int^{\infty}_0  d\omega 
\frac{\omega^4}{2\sinh^2(\omega/2T)}
F(\omega)
\end{equation}
where
\bea
F(\omega)=\nu_{\rm F}^2\int_{\omega/\varv_\mathrm{F}}^\infty\frac{dq}{q} 
|U(q,\omega)|^2.\label{Fomega}
\eea

Due to the $1/\sinh(\omega/2T)$ factor, the integral over $\omega$ is dominated by the region $\omega\sim T$. In the lower-temperature limit, $T\sim\omega\ll \vf\kappa$, 
one can put $\omega=0$ in the function $F(\omega)$,  which reproduces the $T^4\ln T$ scaling of the conductivity, obtained for a statically screened Coulomb potential.
To analyze the opposite limit of  $T
\sim\omega\gg\vf \kappa$, it is convenient to introduce dimensionless variables 
$x\equiv \varv_\mathrm{F}q/\omega$ and $
y=\kappa\varv_\mathrm{F}/\omega$, In terms of these variables, 
\begin{equation}
    F(\omega)=y^2
    \int_1^\infty
    \frac{dx}{x} \frac{x^2-1}{(x+
    y)^2(x^2-1)+
    y^2}\label{F}
\end{equation}
The condition $\omega\gg\vf \kappa$ corresponds to 
$y\ll1$. In this limit, one can neglect $y$ compared  in the denominator of Eq.~\eqref{F}, upon which it is reduced to 
\begin{equation}\label{eqn:int}
     F(\omega)=
     y^2\int_1^\infty
    \frac{dx}{x^3}=\frac{\vf^2\kappa^2}{2\omega^2}
\end{equation}

The final step is to perform the integration over $\omega$ in Eq.~(\ref{eqn:hall-dyn}). 
Upon substituting Eq.(\ref{eqn:int}) into  Eq.~(\ref{eqn:hall-dyn}), the $\omega^2$ in the denominator cancels two powers of $\omega$ from the numerator, leaving us with an overall $\omega^2$ in the integrand, and hence with the $T^2$ scaling of the conductivity.
The final expression 
can be written the form $\delta\sigma^{(\rm ee,0)}_{xx}=-\sigma_{\rm ei} \tau_{\rm i}/\tau_{{\rm ee}}$, where $\sigma_\mathrm{ei}$, which defines  
the effective current relaxation time,
$
\tau_\mathrm{ee}$, 
as given by 
Eq.~(\ref{eqn:tau-dyn}).

In 3D, the polarization bubble for $|\omega|/\vf\leq q\ll \kf$ is given by
\bea
\Pi(q,\omega)=-\nu_{\rm F}\left[1-\frac{\omega}{2\vf q}\ln\frac{\vf q+\omega}{\vf q-\omega}\right]-i\nu_{\rm F}\frac{\pi \omega}{2\vf q}.
\eea
In the limit of $T\sim \omega\gg \vf q$, the dynamic interaction is reduced to the bare Coulomb potential. Accordingly, Eq.~\eqref{Fomega} is replaced by
\bea
F(\omega)\propto \int^\infty_{|\omega|\vf} \frac{dq}{q^4}\propto \frac{1}{\omega^3},
\eea
which gives $\delta\sigma_{xx}^{(\rm ee,0)}\propto T$.
\bibliography{dm_references}
\end{document}
%